\documentclass[a4paper,fleqn,usenatbib]{mnras}

\usepackage{newtxtext,newtxmath}

\usepackage[T1]{fontenc}
\usepackage{multicol}

\def\kms{$\mathrm{km\,s}^{-1}$}

\def\ltsima{$\; \buildrel < \over \sim \;$}
\def\simlt{\lower.5ex\hbox{\ltsima}}
\def\gtsima{$\; \buildrel > \over \sim \;$}
\def\simgt{\lower.5ex\hbox{\gtsima}}

\def\ergsc{{erg cm s$^{-1}$}}
\def\ergs{{erg s$^{-1}$}}
\def\cm2{{cm$^{-2}$}}

\def\xmm{{XMM-{\em Newton}}}
\def\chandra~{{\em Chandra}}

\def\nustar{{\em NuSTAR}}
\def\swift{{\em Swift}}

\def\chandra{{\em Chandra}}

\def\nh{{N$_{\rm H}$}}

\def\f14{{10$^{-14}$}}
\def\f13{{10$^{-13}$}}
\def\f12{{10$^{-12}$}}
\def\f11{{10$^{-11}$}}
\def\e22{{10$^{22}$}}

\def\j1038{{SDSS~J1038+3921}}

\usepackage{graphicx}	
\usepackage{amsmath}	
\usepackage{amssymb}	

\title[Dual AGN systems]{Disclosing the properties of low-redshift dual AGN through XMM-Newton and SDSS spectroscopy}

\author[A. De Rosa et al.]{
Alessandra De Rosa,$^{1}$\thanks{E-mail: alessandra.derosa@iaps.inaf.it.}
Cristian Vignali,$^{2,3}$
Bernd Husemann$^{4}$,
Stefano Bianchi$^{5}$, 
\newauthor
Tamara Bogdanovi\'c$^{6}$,
Matteo Guainazzi$^{7}$ ,
Ruben Herrero-Illana$^{8}$,
S. Komossa$^{9}$, 
\newauthor
Emma Kun$^{10}$, 
Nora Loiseau$^{11}$, 
Zsolt Paragi$^{12}$, 
Miguel Perez-Torres$^{13}$, 
Enrico Piconcelli$^{14}$
\\
$^{1}$ INAF - Istituto di Astrofisica e Planetologia Spaziali (IAPS), via Fosso del Cavaliere, Roma, I-133, Italy\\
$^{2}$ Dipartimento di Fisica e Astronomia, Alma Mater Studiorum, Universit\`a degli Studi di Bologna, Via Gobetti 93/2, 40129 Bologna, Italy \\
$^{3}$ INAF - Osservatorio di Astrofisica e Scienza dello Spazio di Bologna, Via Gobetti 93/3, 40129 Bologna, Italy\\
$^{4}$ Max-Planck Institut f\"{u}r Astronomie, K\"{o}nigstuhl 17, D-69117 Heidelberg, Germany\\
$^{5}$ Dipartimento di Matematica e Fisica, Universit\`a degli Studi Roma Tre, via della Vasca Navale 84, 00146 Roma, Italy\\
$^{6}$ Center for Relativistic Astrophysics, Georgia Institute of Technology, Atlanta, GA 30332, USA\\
$^{7}$ ESA - European Space Research and Technology Centre (ESTEC), Keplerlaan 1, 2201AZ Noordwijk, the Netherlands \\
$^{8}$ European Southern Observatory, Alonso de C\'ordova 3107, Vitacura, Santiago de Chile\\
$^{9}$ Max-Planck-Institut f{\"u}r Radioastronomie, Auf dem H{\"u}gel 69, 53121 Bonn, Germany\\
$^{10}$ Department of Experimental Physics, University of Szeged, D\'om t\'er 9, H-6720 Szeged, Hungary\\
$^{11}$ ESA - European Space Astronomy Centre (ESAC), E-28692 Villanueva de la Ca\~{n}ada, Madrid, Spain\\
$^{12}$ Joint Institute for VLBI ERIC, Postbus 2, NL-7900 AA Dwingeloo, The Netherlands\\
$^{13}$ Centro de Estudios de la F\'{\i}sica del Cosmos de Arag\'on (CEFCA), 44001 Teruel, Spain\\
$^{14}$ INAF - Osservatorio Astronomico di Roma, via Frascati 33, 00040 Monte Porzio Catone (Roma), Italy 
}

\date{Accepted XXX. Received YYY; in original form ZZZ}

\pubyear{2018}

\begin{document}
\label{firstpage}
\pagerange{\pageref{firstpage}--\pageref{lastpage}}
\maketitle

\begin{abstract}
We report on an optical (SDSS) and X-ray (\xmm) study of an optically selected sample of four dual AGN systems at projected separations of 30--60~kpc. 
All sources are detected in the X-ray  band (0.3-10 keV); seven objects are optically identified as Seyfert, while one source, optically classified as a LINER, is likely powered by accretion in virtue of its relatively high X-ray luminosity (1.2$\times10^{41}$~\ergs).  
Six of the eight objects are obscured in X-rays with N$_{\rm H} \geq$ 10$^{23}$ \cm2; three of these, whose X-ray spectrum is dominated by a reflection component, are likely Compton-thick (N$_{\rm H} \geq$ 10$^{24}$ \cm2). 
This finding is in agreement with the hypothesis that galaxy encounters are effective in driving gas inflow toward the nuclear region, thus increasing the obscuration. We compare the absorption properties in our dual AGN with those  in larger samples observed in X-rays but selected in different ways (optical, IR and hard X-rays). We find that the obscured (N$_{\rm H} \geq$ 10$^{22}$ \cm2) AGN fraction within the larger sample is 84$\pm$4 per cent  (taking into account the 90 per cent error on the N$_{\rm H}$ measure) up to large pair separations ($\sim$100~kpc). This is statistically higher than the fraction of obscured AGN in isolated galaxies found in X-ray surveys. 
\end{abstract}




\begin{table*}
\caption{Properties of the sources detected in the \xmm\ observations. \label{tab:obs}}
\begin{tabular}{cccccccccccc}
\hline
Name & srcID &  $z$ & $d$ & FIRST & date  & R$_{\rm extr}$ & pn CR & MOS12 CR & Net exposure \\
SDSS & & & ($''$/kpc) & (mJy) &  & (arcsec) & (10$^{-3}$ s$^{-1}$) & (10$^{-3}$ s$^{-1}$) & pn/MOS12 (ks) \\
(1) & (2) & (3) & (4) & (5) & (6) & (7)  & (8) & (9) & (10) \\
\hline
J094554.41+423840.0  &  src1 & 0.0749 & 21/30 & <0.42 & 27-Apr-2017 & 15 & 532$\pm$5  &  116$\pm$2 & 22/52 \\
J094554.49+423818.7 &  src2 & 0.0752  &  & 1.97$\pm$0.14 &  & 11 & 29$\pm$1  & 6.0$\pm$0.4 & 22/52 \\
\hline
J103853.28+392151.1  & src1 & 0.0551 & 40/43 & 1.19$\pm$0.15 & 04-May-2016 & 25 & 254$\pm{4}$ &66$\pm{1}$ & 18/43 \\
J103855.94+392157.5 & src2 & 0.0548 &  & <0.45 & & 20 & 8.7$\pm{0.7}$&1.5$\pm{0.2}$ & 18/43\\
\hline\
J162640.93+142243.6  & src1 & 0.0482 & 54/51 & 1.85$\pm$0.14 & 26-Aug-2016 & 25 &4.2$\pm{0.6}$ &1.0$\pm{0.2}$ & 53/122 \\
J162644.51+142250.7 & src2  & 0.0487  &  & <0.42 &  & 25 & 1.7$\pm{0.7}$  &0.5$\pm{0.1}$ & 53/122 \\
\hline\
J145627.40+211956.0  & src1 & 0.0446 & 68/59 & <0.45 & 12-Jan-2017 & 18 & 6.0$\pm{0.4}$ &0.9$\pm{0.1}$ & 72/153\\
J145631.36+212030.1 & src2  & 0.0442  &  &1.87$\pm$0.15 & & 18 & 4.3$\pm{0.4}$ &0.7$\pm{0.1}$ & 72/153 \\
\hline
\end{tabular}\\
Col. (1) Name of the source; (2) source ID in each pair (src1 is the one with the highest 2--10~keV  flux); (3) redshift from the stellar continuum in obscured AGN and from narrow emission lines in unobscured AGN (see Sect.~\ref{ssec:optical} for details); (4) angular and projected distance between the sources in each system; (5) 1.4\,GHz flux densities and 3$\sigma$ upper limits from the FIRST survey \citep{beckeretal95}; (6) date of the XMM-Newton observations; (7) radius of the circular region used to extract the XMM-pn and MOS spectra; (8) EPIC-pn and (9) combined MOS count rates in the broadband (0.3--10 keV); (10) EPIC-pn and combined MOS net exposure after filtering correction.
\end{table*}

\section{Introduction}

\label{sec:intro}

Supermassive Black Holes (SMBHs,  with mass 10$^6$-10$^9$ M$_\odot$) are ubiquitous in ellipticals and in the bulges of disk galaxies.
They are likely to affect the evolution of their host galaxy over cosmological time-scales, as suggested by the tight correlation between black hole mass and, {\it e.g.} the bulge stellar velocity dispersion \citep{ferrarese&merritt00,gebhardtetal2000}. The close connection between the formation and evolution of galaxies and of their central SMBHs involves a variety of physical phenomena of great relevance in modern astrophysics \citep{dimatteoetal05,silk&rees98}. 
There is growing evidence that galaxy mergers are the way through which SMBHs can form and evolve, especially at the highest luminosities \citep{treisteretal12}. Numerical simulations have shown that strong inflows in a galaxy merger feed gas to the SMBH, thus powering accretion and triggering the AGN  (e.g. \citealt{dimatteoetal05}).  
Several observational campaigns in different wavebands have also demonstrated that the fraction of dual AGN\footnote{In this paper we define dual AGN all galaxy systems containing two AGN with a separation in the range 0.1--100~kpc.} is higher in galaxies with a close companion, suggesting that galaxy interaction plays a role in AGN triggering \citep{ellisonetal11,satyapaletal14,kossetal12,silvermanetal11,kocevskietal15}. However, other studies found no evidence for an increased AGN fraction in mergers compared to inactive galaxies \citep[e.g.][]{Cisternas:2011,Mechtley:2016}.

The detection and characterisation of dual AGN at kpc scale is fundamental to understand the BH accretion history.
Moreover, dual AGN are the precursor of  coalescing binary SMBHs, which are  strong emitters of gravitational waves \citep{abbottetal16}. 

The search for AGN pairs has received great attention in the last decade (e.g. \citealt{bogdanovic15,komossa&zensus15}), and different methods have been proposed to identify the candidates, depending on their spatial separation. Most of the AGN pairs with 1 to 100\,kpc separations have been identified through extensive optical  \citep{comerfordetal12,liuetal11}, radio \citep{fuetal2015,mullersanchezetal2015}, mid-infrared (mid-IR) \citep{satyapaletal14} and hard X-ray \citep{kossetal10} surveys. 
Recent observations have also demonstrated that dual AGN are characterised by enhanced obscuration with respect to isolated AGN in a parent population \citep{kocevskietal15,satyapaletal17,riccietal17}.
The high penetrative power of hard X-rays (above 2 keV) then provides a unique  tool in the hunt for multiple active nuclei in a galaxy, because they are less affected by contamination from host galaxy emission and absorption, and are produced in large amounts only by AGN (e.g., \citealt{komossaetal03,guainazzietal05,bianchietal08,piconcellietal10,kossetal11,kossetal12}).
The main challenge in this type of study is the need for a statistically significant sample of dual/multiple AGN that covers a wide dynamical range in spatial separations. While a number of AGN pair candidates and merging galaxies have been discovered over the past several years, only a handful of these have eventually been confirmed, usually through intense and observationally expensive multiband follow-up campaigns.

In this paper we characterise four dual AGN systems with separations of 30 to 60\,kpc by means of a multi-wavelength study combining X-ray and optical data. 
We will show that, on the one hand, the optical band is able to select dual AGN candidates even in heavily obscured (N$_H \geq$ 10$^{24}$ \cm2) systems; on the other hand, X-rays allow to robustly confirm/assess their AGN nature and properly characterise the properties (e.g., column density, intrinsic nuclear emission) of the pair members. 
Our study aim at contributing to the still poorly explored sample of confirmed dual AGN with tens-of-kpc separation but still below 60 kpc, since their number is relatively limited (about 30 sources, \citealt{liuetal11,kossetal12,riccietal17}); the present  work enlarges that sample with 8 new objects, i.e. by about 30 per cent.

The paper is organized as follows: in $\S$2 we report on the sample selection, while X-ray observations are introduced in $\S$3. The analysis of the optical and X-ray data is carried out in $\S$4. The main results are discussed in $\S$5 and conclusions are reported in $\S$6. Throughout the paper we adopt a concordance cosmology with H$_{0}$ = 70 km\,s$^{-1}$\,Mpc$^{-1}$, $\Omega_\Lambda$= 0.7, $\Omega_M$ = 0.3. Errors and upper limits quoted in the paper correspond to the 90 per cent confidence level, unless noted otherwise.

\begin{figure*}
\centering
 \includegraphics[width=0.8\textwidth]{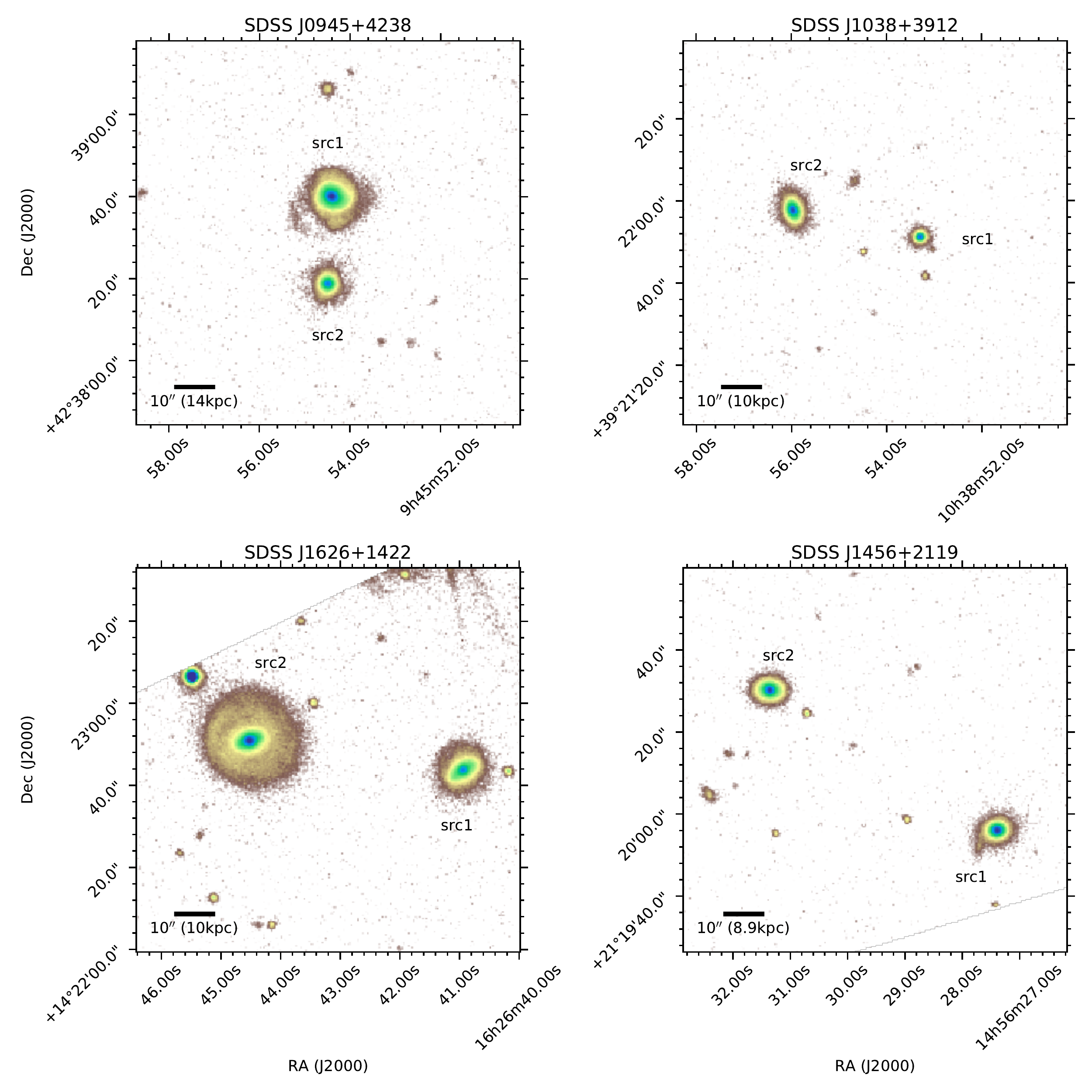}
 \caption{$r$-band SDSS images of our four dual AGN systems shown with a logarithmic surface brightness scaling. The AGN host galaxies are labeled. North is up and East is left. A scale bar indicates $10''$ in all panels.}
\label{fig:SDSS_images}
\end{figure*}

\section{Sample selection}
\label{sec:sample}
We draw our dual AGN sample from an optically selected catalogue of AGN pairs \citep{liuetal11} based on the Sloan Digital Sky Survey (SDSS) Data Release 7 (DR7).  
We then obtained a ``master sample" consisting of 16 systems of Seyfert pairs, 15  with projected separation ranging between 3--60~kpc (2--80~arcsec) and a multiplet \citep{derosaetal15}, with a redshift distribution $z=0.03-0.17$. 

The systems with angular separation larger than 10~arcsec have been proposed to be observed with \xmm, while closest pairs (angular separation lower than 10~arcsec) will be proposed for \chandra\ observations. Four of the larger-separation dual AGN systems are studied in this work (see Table~\ref{tab:obs}); they have  projected separations ranging from 30 to 59~kpc. In Fig.~\ref{fig:SDSS_images} we show the SDSS $r$-band image for all our four galaxy pairs and label them in agreement with the nomenclature in Tab. \ref{tab:obs}.

\begin{figure*}
\centering
\vglue-2cm
\includegraphics[width=0.53\textwidth,angle=-90]{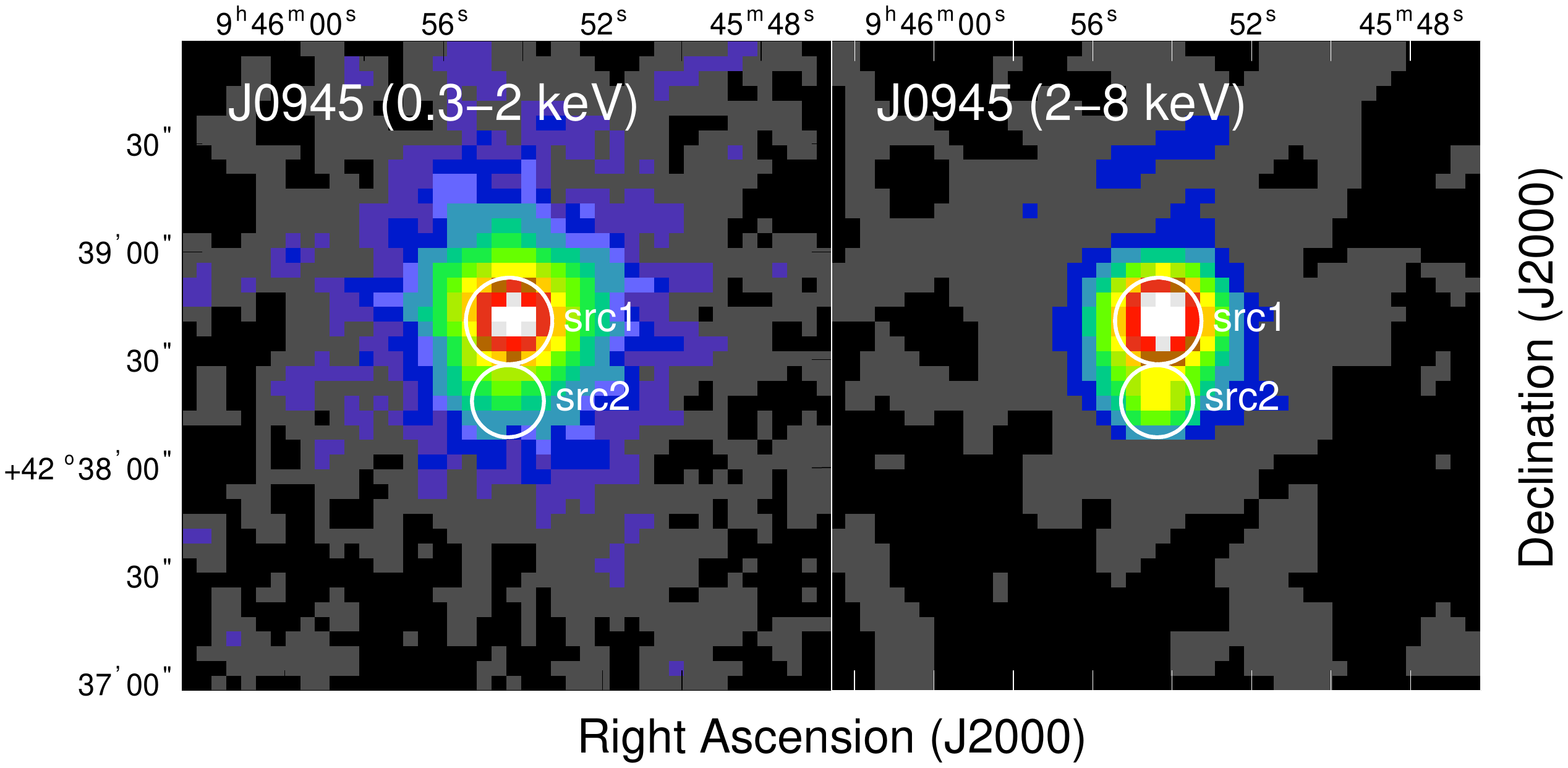}
\vglue-3.9cm
\includegraphics[width=0.53\textwidth,angle=-90]{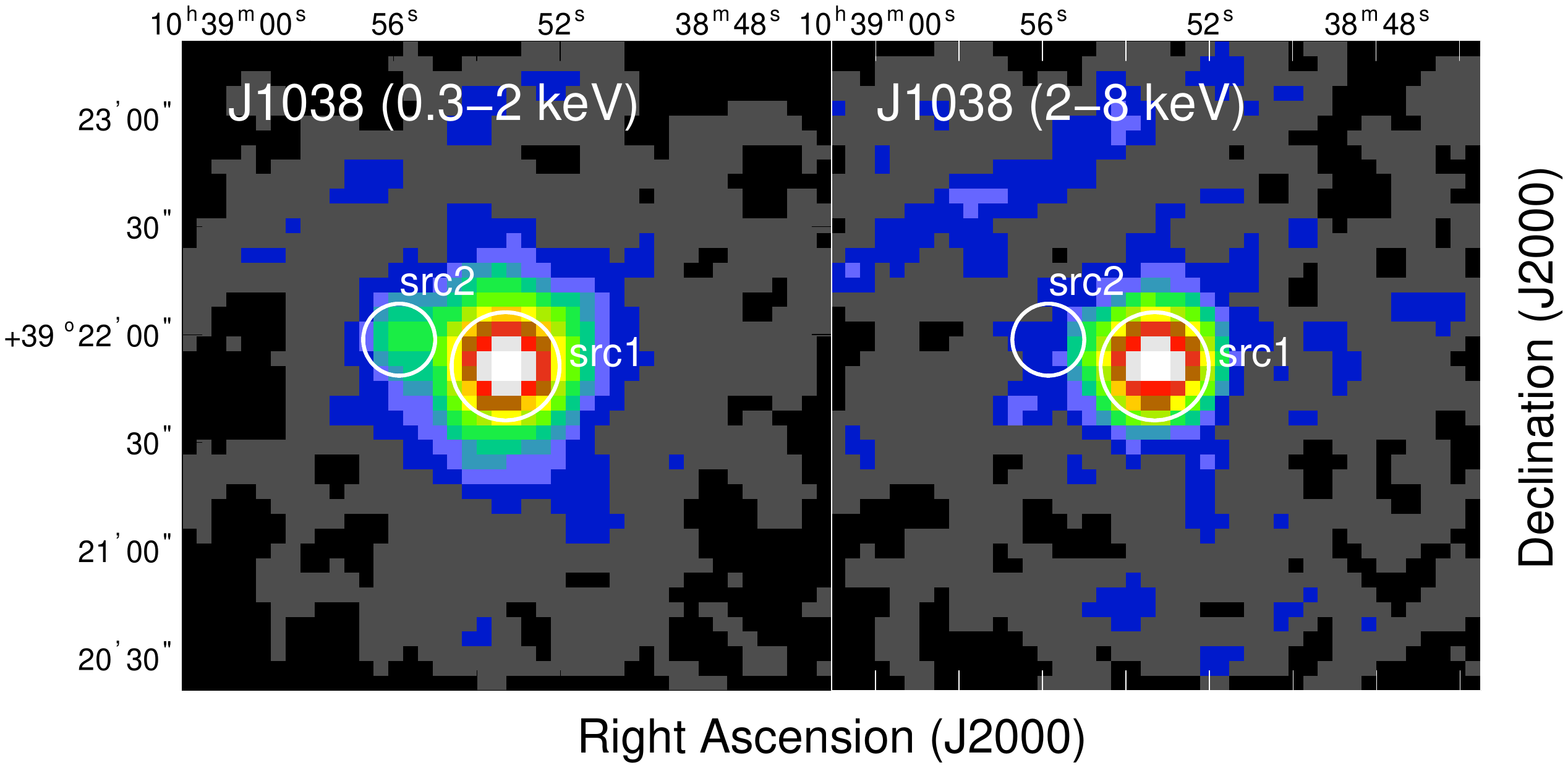}
\vglue-3.9cm
\includegraphics[width=0.53\textwidth,angle=-90]{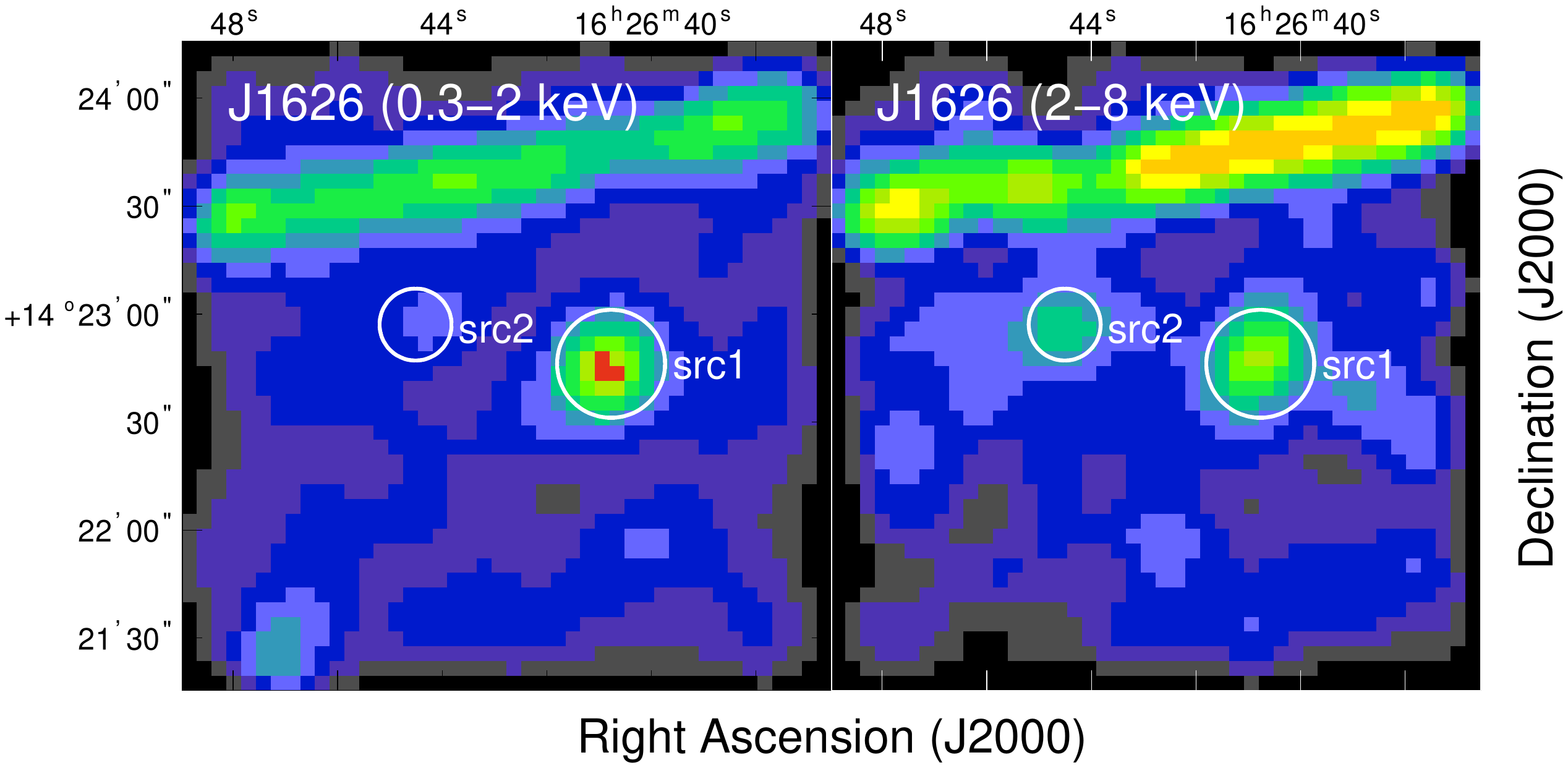}
\vglue-3.9cm
\includegraphics[width=0.53\textwidth,angle=-90]{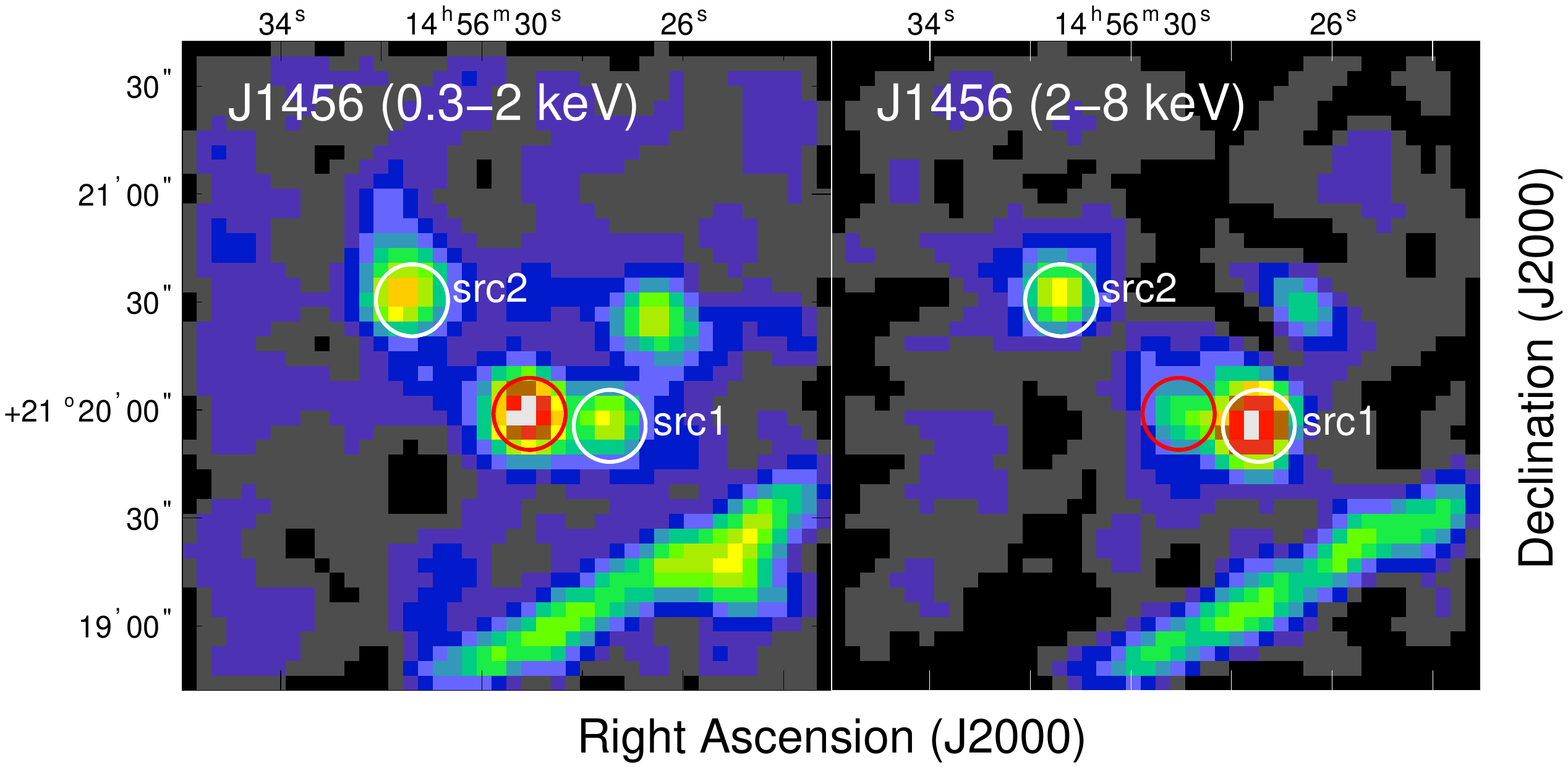}
\vglue-1.9cm
\caption{\xmm-pn 3$\times$3 arcmin$^{2}$ region around the systems investigated in this paper. 
Images in the 0.3--2~keV and 2--8~keV band are shown in the left and right panels, respectively. From top to bottom: SDSS J0945+4238, SDSS J1038+3921, SDSS J1626+1422, SDSS J1456+2119. The dual AGN (where src1 it the pair member with the higher 2--10~keV flux) are marked by a solid circle (whose size is not representative of the source extraction region reported in Tab.\ref{tab:obs}). North is up and East to the left. Bright elongated regions in J1626 and J1456 images, due the CCD edges in the pn camera, did not affect our data analysis. The soft X-ray bright excess between src1 and src2 in J1456 field (marked with a red circle) is a background source and is discussed in Sect.\ref{subsect: x-ray}}
\label{fig:xmm_reg}
\end{figure*}

\section{X-ray observations} 
\label{ssec:xmm_obs} 
Both EPIC cameras (pn and MOS) were observing in full-frame mode and with the thin filter applied. Data were reduced using SAS v13.5 with standard settings and the most updated calibration files available at the time of the data reduction. Appropriate filtering to remove periods of high and flaring background was applied; the net exposure for the EPIC cameras is reported in the last column of Table~\ref{tab:obs}. 

The EPIC-pn images of the systems are shown in Fig.~\ref{fig:xmm_reg} in two different energy ranges, 0.3--2~keV (left) and 2--8~keV (right). Sources in the \xmm\ fields were detected using the EPIC source detection tool {\tt edetect\_chain} in five energy ranges ($0.3-0.5$~keV, $0.5-1$~keV, $1-2$~keV, $2-4.5$~keV, $4.5-12$~keV), adopting a threshold of 3$\sigma$. All spectra were extracted from circular regions with radii in the range 11$''$--25$''$ (see Table \ref{tab:obs}), depending on the source counts and the separation of the two sources in each system; these regions include $\sim$60--90\% of the source counts at 1.5~keV in the EPIC cameras. Background spectra were extracted in the same CCD chip using circular regions free from contaminating sources.

Spectra were rebinned in order to have at least 20 counts for each background-subtracted spectral channel and not to oversample the intrinsic energy resolution by a factor larger than 3. In the following we indicate as src1 (src2) the source with the highest (lowest) hard X-ray flux in each pair. 
Spectral fits for pn and (co-added) MOS cameras were performed in the 0.3--10~keV energy band.

\begin{table*}
\caption{Optical emission lines parameters from the SDSS spectra for the narrow line components.}
\label{tab:sdss}
\begin{small}
\begin{tabular}{ccccccccccc}\hline\hline
SrcID & H$\beta$ & [OIII]\,$\lambda$5007 & [OI]\,$\lambda$6300 & H$\alpha$ & [NII]\,$\lambda$6583 & [SII]\,$\lambda$6717,30 & $\sigma_\mathrm{narrow}$ & $\sigma_\mathrm{broad}$ & class\\
& \multicolumn{6}{c}{Flux [10$^{-16}$ \ergsc]} & \multicolumn{2}{c}{[\kms]} & \\\hline
\multicolumn{10}{c}{SDSS J0945+4238}\\\hline
src1 & $14.3\pm2.4$ & $70.2\pm3.2$ & $3.7\pm1.0$ & $61.4\pm6.7$ & $32.7\pm5.2$ & $19.1\pm1.6$ & $101\pm9$ & $243\pm13$ & AGN1\\
src2 & $47.4\pm0.1$ & $201.8\pm0.1$ & $14.2\pm0.2$ & $236.4\pm0.1$ & $165.2\pm0.1$ & $89.3\pm0.4$ & $143\pm1$ & ... & AGN2\\\hline
\multicolumn{10}{c}{SDSS J1038+3921}\\\hline
src1 & $8.2\pm1.0$ & $47.8\pm1.0$ & $18.5\pm0.7$ & $32.7\pm2.0$ & $32.7\pm1.4$ & $33.5\pm1.0$ & $160\pm3$ & ... & AGN1\\
src2 & $12.2\pm0.1$ & $125.7\pm0.1$ & $6.1\pm0.2$ & $47.1\pm0.1$ & $32.7\pm0.1$ & $24.6\pm0.5$ & $101\pm1$ & ... & AGN2\\\hline
\multicolumn{10}{c}{SDSS J1626+1422}\\\hline
src1 & $16.5\pm0.3$ & $179.1\pm0.6$ & $17.4\pm0.8$ & $73.8\pm0.7$ & $72.8\pm0.6$ & $45.6\pm1.4$ & $83\pm1$ & $289\pm3$ & AGN2\\
src2 & $2.9\pm0.2$ & $3.8\pm0.1$ & $1.1\pm0.2$ & $7.1\pm0.3$ & $10.2\pm0.3$ & $6.6\pm0.6$ & $129\pm10$ & ... & LINER \\\hline
\multicolumn{10}{c}{SDSS~J1456+2119}\\\hline
src1 & $5.3\pm0.6$ & $40.1\pm1.6$ & $2.7\pm0.5$ & $21.5\pm1.4$ & $13.3\pm0.9$ & $7.8\pm1.3$ & $75\pm2$ & $184\pm7$ & AGN2\\
src2 & $16.5\pm0.6$ & $203.0\pm0.5$ & $16.9\pm0.6$ & $80.9\pm1.5$ & $72.6\pm1.6$ & $32.5\pm1.2$ & $97\pm1$ & $450\pm2$ & AGN2\\\hline
\end{tabular}\end{small}

\end{table*} 

\section{Data analysis and Results} 
\label{sec:data_analysis}
\subsection{SDSS optical spectra} 
\label{ssec:optical}
We retrieved the SDSS DR12 spectra \citep{SDSS-DR12} for all sources listed in Table~\ref{tab:obs} from the survey webpage\footnote{http://skyserver.sdss3.org/dr12.}. All  spectra were taken at the location of the galaxies as shown in Fig.~\ref{fig:SDSS_images} and are reported in the Appendix \ref{sec:appendix}.
We analyse the spectra mainly to infer the narrow emission-line flux for primary diagnostic lines such as H$\beta$, [OIII] $\lambda5007$, [OI] $\lambda6300$, H$\alpha$, [NII] $\lambda6583$ and [SII] $\lambda\lambda6717,6730$. The analysis of Seyfert 1 and Seyfert 2 galaxy spectra varies significantly so we explain both approaches in the following.

\subsubsection{Emission-line modelling of obscured AGN}
The light of the nucleus is blocked in obscured AGN. Only the stellar light of the host galaxy and the ionized gas from the narrow-line region (NLR) and/or star forming region can be seen. We measure the emission-line fluxes on top of the stellar continuum light with \textsc{PyParadise} software \citep[e.g.][]{Husemannetal16,Weaveretal18}. \textsc{PyParadise} models the stellar continuum as a superposition of template stellar spectra from the INDO-US stellar library \citep{Valdesetal04} after normalizing both the SDSS and the template spectra with a running mean of 100\,pixel, interpolating regions with strong emission lines. A simple Gaussian kernel  is used to match the template spectra to the line-of-sight velocity distribution. 
\begin{figure*}
 \includegraphics[width=\textwidth]{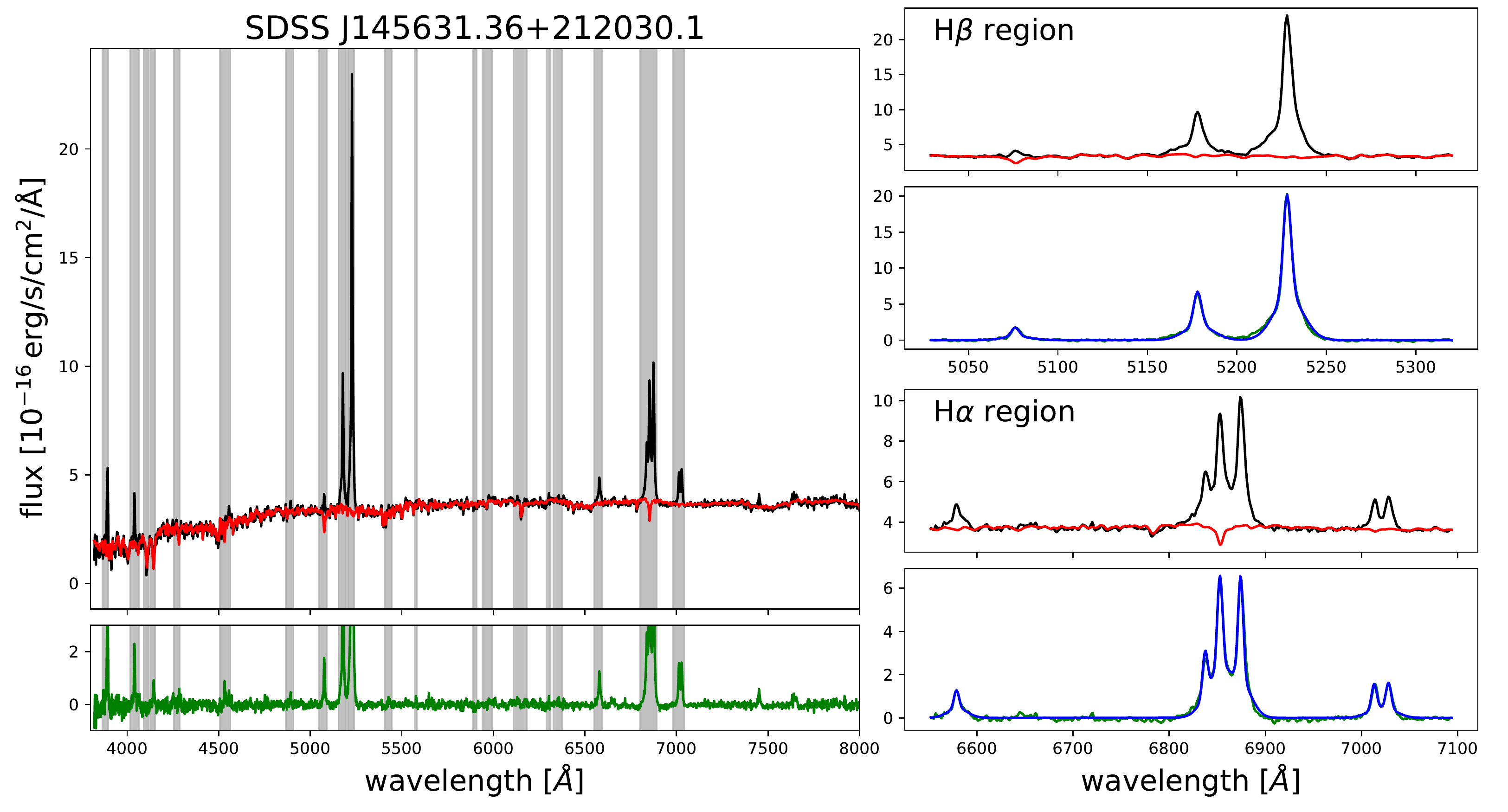}
 \caption{Continuum and emission-line modelling for SDSS~J145631.36+212030.1 as an example. \textit{Left panel:} SDSS spectrum is shown as the black line with best-fit stellar continuum from PyParadise shown in red. The residual spectrum is shown below in green color. The gray shaded area highlight regions that were masked during the model because they contain strong emission lines, atmospheric absorption bands, or strong sky-line residuals. \textit{Right panels:} Zoom into the wavelength region focussed around H$\beta$ (upper spectra) and H$\alpha$ (lower spectra). Again the data is shown in black and the stellar continuum model in red. Below the best-fit emission-line model is shown in blue on top of the green residual spectrum. Two Gaussian components are clearly needed to provide a good model to the emission lines in this case.
}
 \label{fig:type2_fitting}
\end{figure*}

The emission-line fluxes are modelled in the residual spectrum after the best-fit continuum model is subtracted (see Fig.~\ref{fig:type2_fitting}). We couple all Gaussian line profiles in redshift and intrinsic rest-frame velocity dispersion which reduced the degree-of-freedom and makes the fit robust even for faint emission lines. In some cases a second kinematics component is seen in the [OIII] doublet and other forbidden lines which we add as a second system of coupled Gaussians. Errors are obtained using a bootstrap approach where 100 realizations of the spectrum were generated based on the pixel errors with just 80\% of the template spectra and emission-lines are modelled again with the same approach (at fixed stellar kinematics). The resulting emission-line parameters are listed in Table~\ref{tab:sdss} together with one or two velocity dispersions if required for the respective model. As shown in Fig.~\ref{fig:type2_fitting} we detect a broad component in the [OIII] line of SDSS~J145631.36+212030.1 with a FWHM of about 1000\,km/s. This indicates a powerful ionized gas outflow in this Compton-thick (CT) AGN.

\subsubsection{Emission-line modelling of unobscured AGN}
\begin{figure}
 \includegraphics[width=\columnwidth]{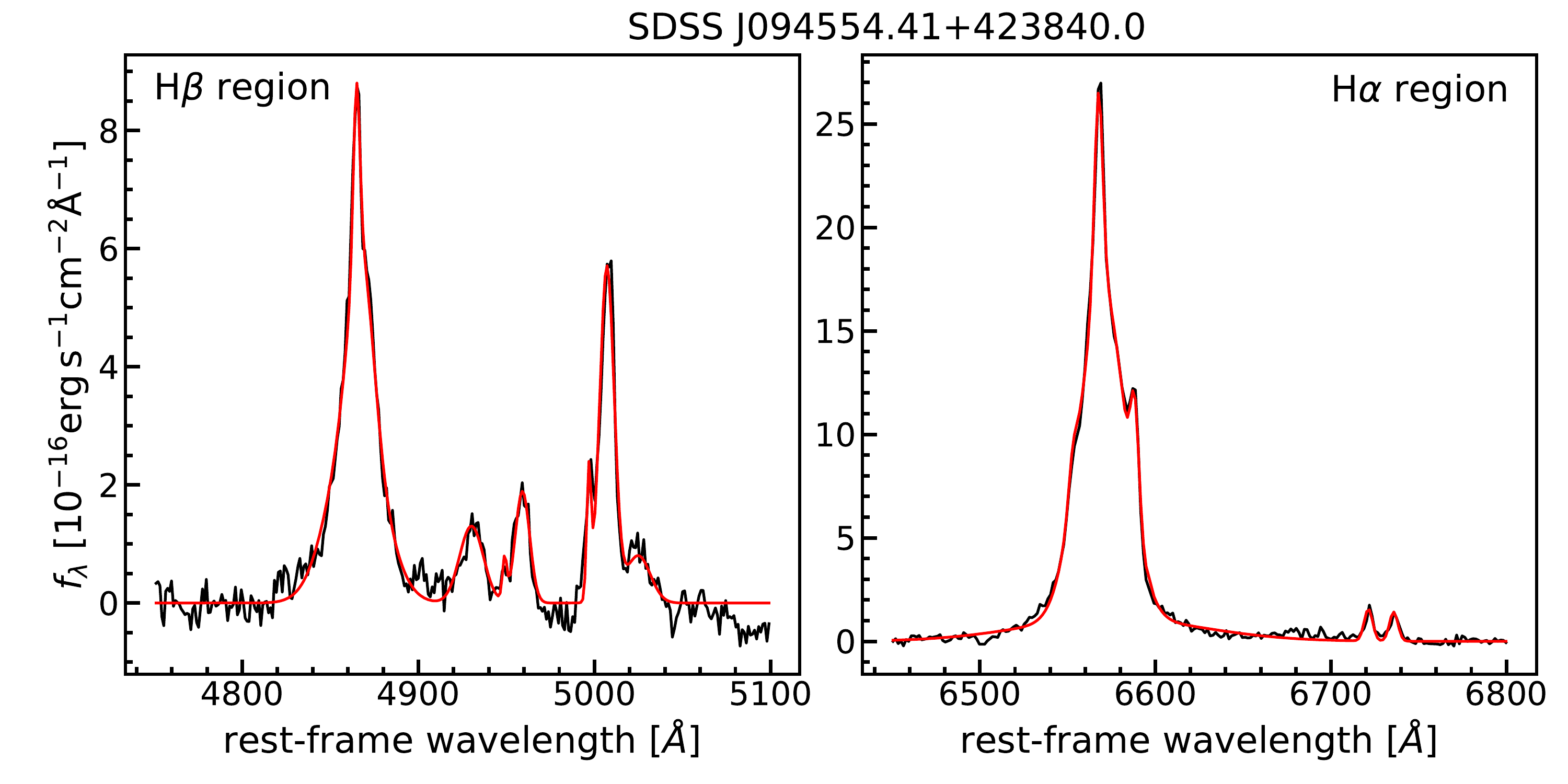}\\
 \includegraphics[width=\columnwidth]{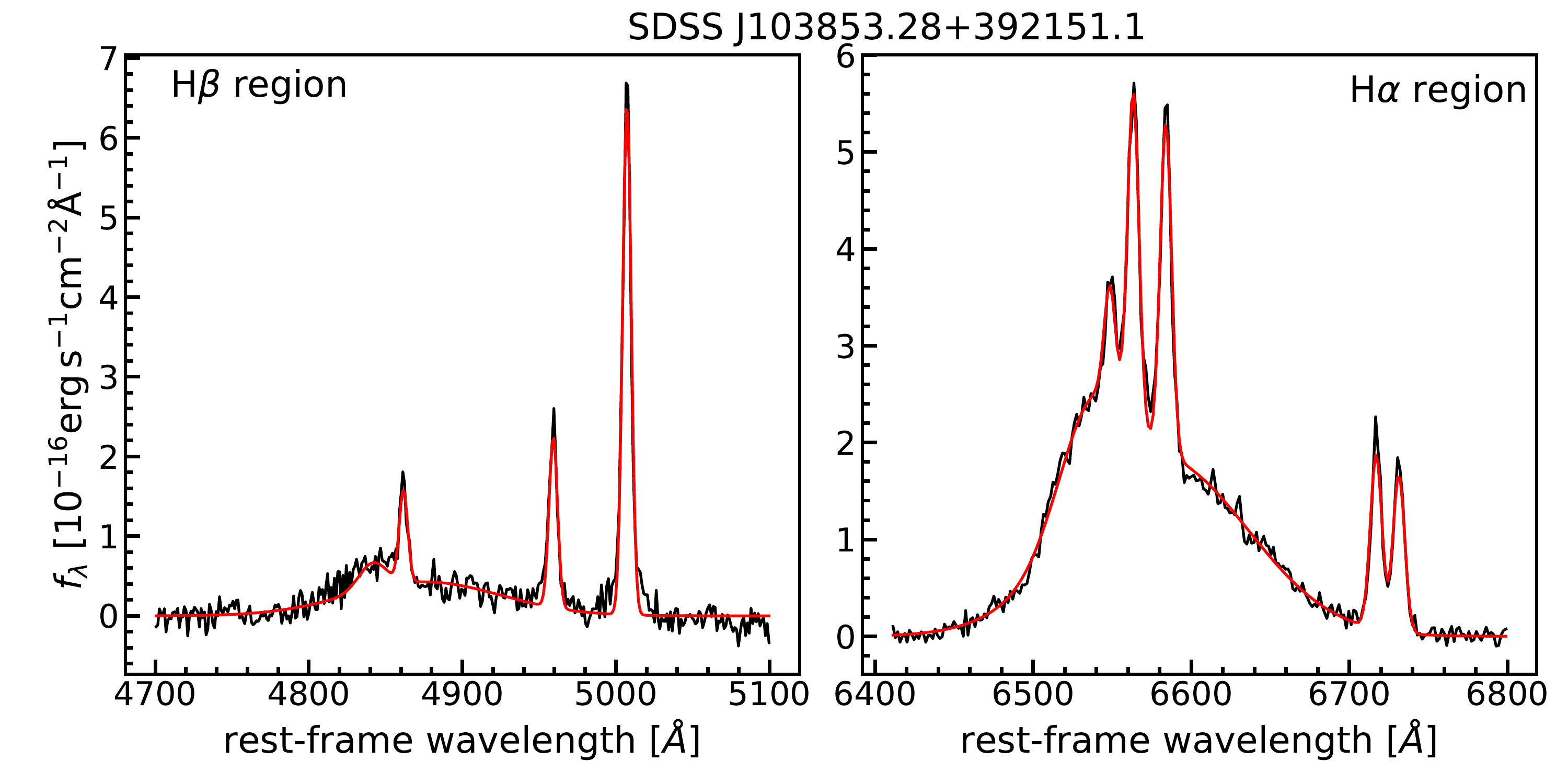}
 \caption{Emission line modeling of the two unobscured type 1 AGN spectra. J0945 in shown in the upper panels with strong Fe\,II doublet while J1038 is shown in the bottom panels. The continuum is approximated by a local linear relation and the broad H$\beta$ and H$\alpha$ lines are independently fitted by 2 or 3 broad Gaussian components. The narrow lines are coupled in intrinsic redshift and velocity dispersion for one or two systems of Gaussians.}
 \label{fig:type1_fitting}
\end{figure}

The two unobscured AGN among our dual AGN systems need to be modelled differently as a power law continuum with broad Balmer lines contribute on top of the stellar continuum and narrow lines. Although the stellar continuum still contributes to the spectrum, we ignore it for our emission-line purposes as we focus here on the broad Balmer line measurements, particularly H$\alpha$, to estimate the BH mass \citep[e.g.][]{Greene:2007}. First, we subtract a local pseudo-continuum from H$\beta$ and H$\alpha$ emission-line regions, which we approximate as a linear function based on the emission-line free regions left and right of the H$\beta$-[OIII] and the H$\alpha$-[NII]-[SII] region, respectively. Then, we model the broad Balmer lines with two kinematically independent Gaussians and a set of coupled Gaussians for the narrow emission lines. The coupling of the narrow emission lines is essential for a robust modelling of the [NII] doublet and narrow H$\alpha$ which significantly blend with the broad H$\alpha$ component. Due to the typical blue wing asymmetry in the narrow lines of AGN \citep[e.g.][]{Mullaney:2013}, we use two kinematic components coupled for all narrow emission lines. Additionally, strong Fe\,II emission lines can be present in unobscured AGN spectra which are typically modelled using Fe\,II template. However, the Fe\,II template consists only of two Fe\,II lines in the modelled wavelength region which we model as Gaussians kinematically coupled to the broad H$\beta$ line components . This way we avoid Fe\,II template mismatches for our individual targets, but still achieve a very good fit to the spectra (see Fig.~\ref{fig:type1_fitting}). 

The results for the narrow emission lines are included in Table~\ref{tab:sdss} together with the obscured AGN. The broad-line parameters are listed in Table~\ref{tab:type1} instead. The broad H$\alpha$ line of SDSS J094554.41+423840.0 is $1442$~km s$^{-1}$ (FWHM). It is significantly narrower than typical broad-line AGN and therefore is classified as a narrow-line Seyfert 1 (NLSy1). For both AGN we compute the BH masses using the broad H$\alpha$ luminosity and line width based on the calibration of \citet{Greene:2005}. This way we avoid that any host galaxy continuum affacts our BH mass estimates.

\begin{table*}
\caption{Broad H$\alpha$ measurements for the two unobscured AGN.}
\label{tab:type1}
\begin{small}
\begin{tabular}{ccccc}\hline\hline
Object & $f_{\mathrm{H}\alpha}$ & $\log(L_{\mathrm{H}\alpha})$ & $\mathrm{FWHM}_{\mathrm{H}\alpha}$ & $\log(M_\mathrm{BH})$ \\
 & [$10^{-14}\,\mathrm{erg}\,\mathrm{s}^{-1}\,\mathrm{cm}^{-2}$] & [erg\,s$^{-1}$] & [km\,s$^{-1}$] & [$M_\odot$] \\\hline
SDSS~0945$+$4238 & $7.06\pm0.38$ & $41.99\pm0.02$ & $1442\pm75$ & $6.6\pm0.1$  \\
SDSS~1038$+$3921 & $3.24\pm0.17$ & $41.37\pm0.02$ & $5557\pm375$ & $7.5\pm0.1$\\\hline
\end{tabular}\end{small}

\end{table*}

\subsection{Emission-line classification}
\begin{figure*}
\includegraphics[width=\textwidth]{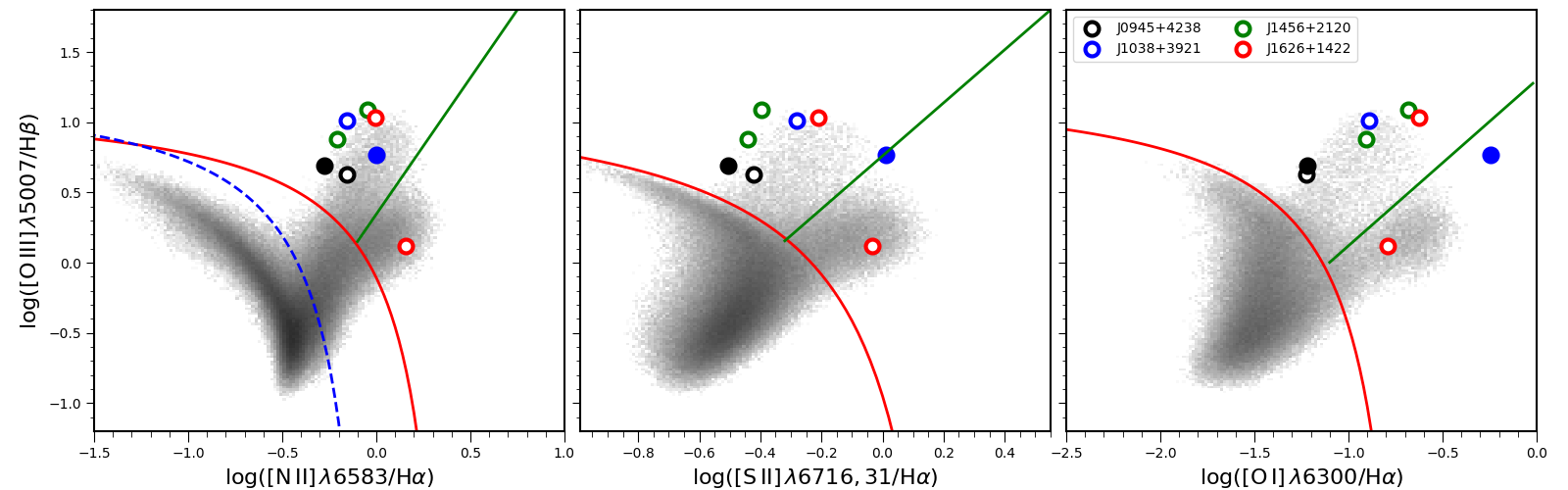}
\caption{Classical emission-line ratio diagnostic diagrams for all spectra of our dual AGN systems. Errors bars for these high signal-to-noise ratio (S/N) spectra are smaller than the symbol size. Obscured AGN are denoted by open circles and unobscured AGN are marked as filled circles. The distribution of line ratios in the overall galaxy population are shown as the gray density map taken from the SDSS DR7 MPA-JHU value-added catalog \citep{Brinchmannetal04}. Red solid line marks the maximum line starburst as inferred by \citealt{Kewleyetal01}, the blue dashed line represents the empirical SFG boundary proposed by \citealt{kauffmannetal03}, and the green solid line is the empirically proposed division between AGN and LINERs \citep{Kewleyetal06,CidFernandesetal10}.} \label{fig:bpt}
\end{figure*}

We use the inferred narrow-line fluxes to construct standard emission-line diagnostic diagrams for the unobscured and obscured AGN to classify the emission in star-forming, LINER-like or AGN-like as shown in Fig.~\ref{fig:bpt}. Standard demarcation lines are used to seperate those ionization sources in the three diagnostics diagrams based on the [OIII]/H$\beta$, [NII]/H$\alpha$, [OI]/H$\alpha$ and [SII]/H$\alpha$ line ratios \citep{Kewleyetal01,kauffmannetal03,Kewleyetal06,CidFernandesetal10}. While such a classification is not necessary for the unobscured AGN as they are already identified as AGN through the presence of broad lines, it is a crucial exercise for the obscured AGN.

We find that all but one of our dual AGN candidates are classified as AGN-like spectra, while one turns out to be a LINER-like spectrum purely based on the emission line ratios. All classifications are consistent in all three diagnostic diagrams which highlight the robustness of the classification in this case. The only exception in our sample is the unobscured AGN SDSS~J103853.28+392151.1. While the [NII]/H$\alpha$ line ratio puts the object into the Seyfert-like classification in the classical BPT, the excess in the [OI]/H$\alpha$ and [SII]/H$\alpha$ leads to a LINER-like classification in the other diagrams. This specific excess  in the narrow [OI] and [SII] lines with respect to [NII] is very unusal in the AGN population.

The most important point from this analysis is that all dual AGN systems are confirmed as such. Only the ionization source of the LINER-like emission still has to be verified by the X-ray data (described later). 

\begin{table*}
\caption{\xmm\ spectral analysis results. 
\label{tab:xmm_analysis}}
\begin{tabular}{ccccccccccccc}
\hline
SrcID & $\Gamma$ &  $\Gamma_{\rm soft}$/kT & N$_{\rm H}$ & f$_{\rm scatt}$ & F$_{\rm soft}^{\rm oss}$ & F$_{\rm hard}^{\rm oss}$ & L$_{\rm x}^{\rm oss}$ & L$_{\rm x}$  & L$_{\rm Bol}$ &  L$_{\rm [OIII]}$& $\chi^2$/dof & X-ray class\\
(1) & (2) & (3) & (4) & (5) & (6) & (7) & (8) & (9) & (10) & (11) & (12) & (13)\\
\hline
\multicolumn{11}{c}{SDSS~J0945+4238}\\\hline
src1  & 2.4$^{+0.1}_{-0.1}$ & 0.11$\pm{0.01}$& $<$0.03 & - & 68$\pm{1}$ & 50$\pm{3}$ & 7.0$\pm{0.3}$  &7.0$\pm{0.3}$ &  112 & 0.2 & 224/231 & U \\
src2 & 1.9$^\star$ &  2.6$\pm{0.2}$ & 24$^{+8}_{-6}$ & 13$\pm$5 & 3.5$\pm{0.2}$ & 19$\pm{2}$ & 2.4$\pm{0.3}$ & 6.9$\pm{0.3}$ &  110 & 1.1 & 22/26  & C-thin\\
\hline
\multicolumn{11}{c}{SDSS~J1038+3921}\\\hline
src1  & 1.62$\pm{0.07}$ &  0.12$^{+0.03}_{-0.03}$ & $<$0.03 & - & 27$\pm{1}$ &55$\pm{5}$ & 4.2$\pm{0.2}$ & 4.2$\pm{0.2}$ & 53 & 0.08 & 140/152 & U \\
src2 & 1.9$^\star$& 3.0$\pm{0.3}$ & $>100$ &  $>$1 & 1.0$\pm{0.1}$ & 1.6$\pm{1.0}$ & 0.2$\pm{0.1}$ & 8$^\dagger$ & 132 & 0.2 & 21/29 & C-thick \\
\hline
\multicolumn{11}{c}{SDSS J1626+1422}\\\hline
src1  & 1.9$^\star$ & 2.2$\pm{0.5}$ & 67$^{+150}_{-50}$& 3$\pm$2 & 0.4$\pm{0.1}$& 2.3$^{+0.8}_{-2.0}$ &  0.2$\pm{0.1}$ & 0.8$^{+0.3}_{-0.7}$ &  8 & 0.3 & 36/39 & C-thick$\dagger\dagger$ \\
src2 & 1.9$^\star$  & =$\Gamma$  & 6$^{+12}_{-4}$ & $<$1 & $<$0.2 & 1.2$\pm{0.8}$ & 0.5$\pm{0.1}$& 0.12$\pm{0.08}$ & 1 & 0.001 & 13/18 & C-thin \\
\hline
\multicolumn{11}{c}{SDSS J1456+2119}\\\hline
src1  & 1.9$\pm{0.9}$ & 2.8$\pm{0.5}$ & 75$^{+28}_{-23}$ & 0.7$\pm$0.4 & 0.28$^{+0.04}_{-0.02}$ & 8.6$^{+0.5}_{-3.1}$ & 0.4$^{+0.1}_{-0.1}$ &3.7$^{+0.2}_{-1.3}$ & 50 & 0.04 & 43/45 & C-thin \\
src2 & 1.9$^\star$ & 2.8$\pm{0.5}$ & $>100$  & 3$\pm$2 & 0.40$\pm{0.05}$ & 2.3$\pm{0.5}$ &0.2$\pm{0.1}$  & 8$^\dagger$ & 132 & 0.3 & 45/37 & C-thick \\
\hline
\end{tabular}

{(1) Source ID in each pair; (2) Primary photon index; (3) Soft X-ray photon index of the scattered power-law (obscured AGN - C-thin/thick) or temperature of the thermal black body component (unobscured AGN - U); (4) Absorption column density in units of 10$^{22}$~cm$^{-2}$;(5) Scattered fraction (\%), for the type 2 AGN, obtained as the ratio between the luminosity of the scattered component and the primary continuum in 0.3--2~keV range (see Sect.\ref{sec:data_analysis}); (6) Observed soft (0.5--2~keV) and (7) hard (2--10~keV) flux in units of $10^{-14}$ erg~cm$^{-2}$~s$^{-1}$; (8) Observed and (9) absorption-corrected, rest-frame 2--10~keV luminosity in units of $10^{42}$~erg~s$^{-1}$. $^\dagger$ For Compton-thick sources, the intrinsic 2--10~keV luminosity has been obtained by multiplying the observed luminosity by 80 \citep{lamastraetal09,marinuccietal12}; (10) Bolometric luminosity in units of $10^{42}$~erg~s$^{-1}$ obtained using \cite{marconietal04} bolometric corrections; (11) De-reddened [O\ III] luminosity in units of $10^{42}$~erg~s$^{-1}$ (see Table~\ref{tab:sdss}); (12) $\chi^2$ over degrees of freedom of the best-fitting model (see Section \ref{sec:data_analysis}.); (13) Classification from the X-ray spectral analysis presented in this work (U: unobscured; C-thin: Compton-thin; C-thick: Compton-thick). $^\star$: Fixed parameter during the fit. $\dagger\dagger$ CT candidate.
}
\end{table*}

\subsection{X-ray spectral analysis}
\label{subsect: x-ray}
All sources in our sample have been detected at a confidence level $\gtrsim3\sigma$ in the full X-ray  (0.3--10 keV)  band. For all sources we fitted the pn and co-added MOS data in the 0.3--10 keV energy band with a baseline model (BLM) composed by (1) an absorbed (\texttt{zPHABS} model in {\sc xspec} with associated cross sections from \citealt{balucinska&mccammon92}) power law that represents the emission of the central regions as due to Comptonization of electrons in a hot-corona by seed photons, probably originated in the accretion disc (the disc-hot corona scenario, see \citealt{haardt&maraschi93,haardtetal94}), (2) a soft, unabsorbed emitting component that reproduces the data at energies lower than $\sim$2~keV, the so called ``soft excess". The physical origin of this component is different in obscured and unobscured AGN.
In obscured AGN it should be due to star formation activity \citep{iwasawaetal11}, scattering of the primary X-ray emission in Compton-thin circumnuclear gas \citep{uedaetal07} or a blend of radiative recombination transitions in a photoionized gas \citep{bianchietal06}. In unobscured AGN the soft excess is often attributed to blurred relativistic reflection \citep{crummyetal06} or Comptonization of the seed optical/UV photons in plasma colder than that responsible for the primary X-ray component; this model is referred to as warm Comptonization (\citealt{petruccietal18}, and references therein).  
To phenomenologically reproduce the soft excess in obscured AGN we used a scattered power-law; the fraction of scattered component is measured by the parameter f$_{\rm scatt}$, that is the ratio between the luminosity of the scattered component vs the primary component in 0.3--2 keV energy range. For unobscured AGN, we reproduced the soft excess with a phenomenological model, a thermal black body emission. For sources with low X-ray photon statistics, the primary power law photon index has been frozen to $\Gamma=1.9$, as expected for AGN (e.g., \citealt{bianchietal09}). The power-law and the black body components are not physical models, but they are able to reproduce the excess through the softer energies, including any possible contribution from the star formation to the luminosity in the $\approx$0.3--2 keV band; a more detailed discussion about the estimate of the star-formation rate (SFR) is presented in Sect. \ref{sec:disc}.

When statistically needed, we included in the fit physically motivated components to fit narrow-band features as emission/absorption lines.  Reprocessing from circumnuclear material, the Compton hump, was taken into account assuming  a semi-infinite optically-thick slab ({\tt PEXRAV} model in {\sc xspec}; \citealt{magdziarz&zdziarski95}). The inclination and the metallicity of the slab were fixed to 45$^\circ$  and to solar, respectively. The reflection fraction was measured using the parameter R = $\Omega$/2$\pi$, where $\Omega$ is the covering factor of the reflecting material.
The whole model is absorbed by Galactic gas with column density obtained by \cite{dickey&lockman90}\footnote{http://heasarc.gsfc.nasa.gov/cgi-bin/Tools/w3nh/w3nh.pl.}.

The luminosity of the ${[\rm OIII]\lambda}$5007  emission line  is considered a good indicator of the intrinsic luminosity of the AGN, in both type~1 and type~2 (see also Sect. \ref{ssec:optical}), once extinction within the NLR and, at some level, its geometry, are properly taken into account. 
In particular, we use the Balmer decrement (H$\alpha$/H$\beta$ ratio) to correct the [O\ III] emission for the extinction. 
To derive the L$_{\rm [OIII]}$ corrected for extinction, we used the relation from \cite{bassanietal99} which assumes the \cite{cardellietal89} extinction law and an intrinsic Balmer decrement equal to 3; this value represents the case for the NLR \citep{osterbrock&ferland06}.
The ratio between the X-ray luminosity and the L$_{\rm [OIII]}$ $\lambda$5007 has been therefore used as an indirect measurement of \nh.
The L$_{\rm x}$/L$_{\rm [OIII]}$ ratio has been investigated by several authors using different samples of type~1 and type~2 AGN \citep{heckmanetal05, mulchaeyetal94, bassanietal99, lamastraetal09, vignalietal10}. 
In particular, \cite{marinuccietal12} measured log~(L$_{\rm x}$/L$_{\rm [OIII]}$)=-0.76  (0.1 dex dispersion) in a sample of Compton-thick AGN, while for Compton-thin AGN \cite{lamastraetal09} found 
log~(L$_{\rm x}$/L$_{\rm [OIII]}$)=1.09  (0.63 dex dispersion).
These values are useful to shed further light on the source classification based on the X-ray spectral analysis.

\begin{figure*}
\includegraphics[scale=.31,angle=-90]{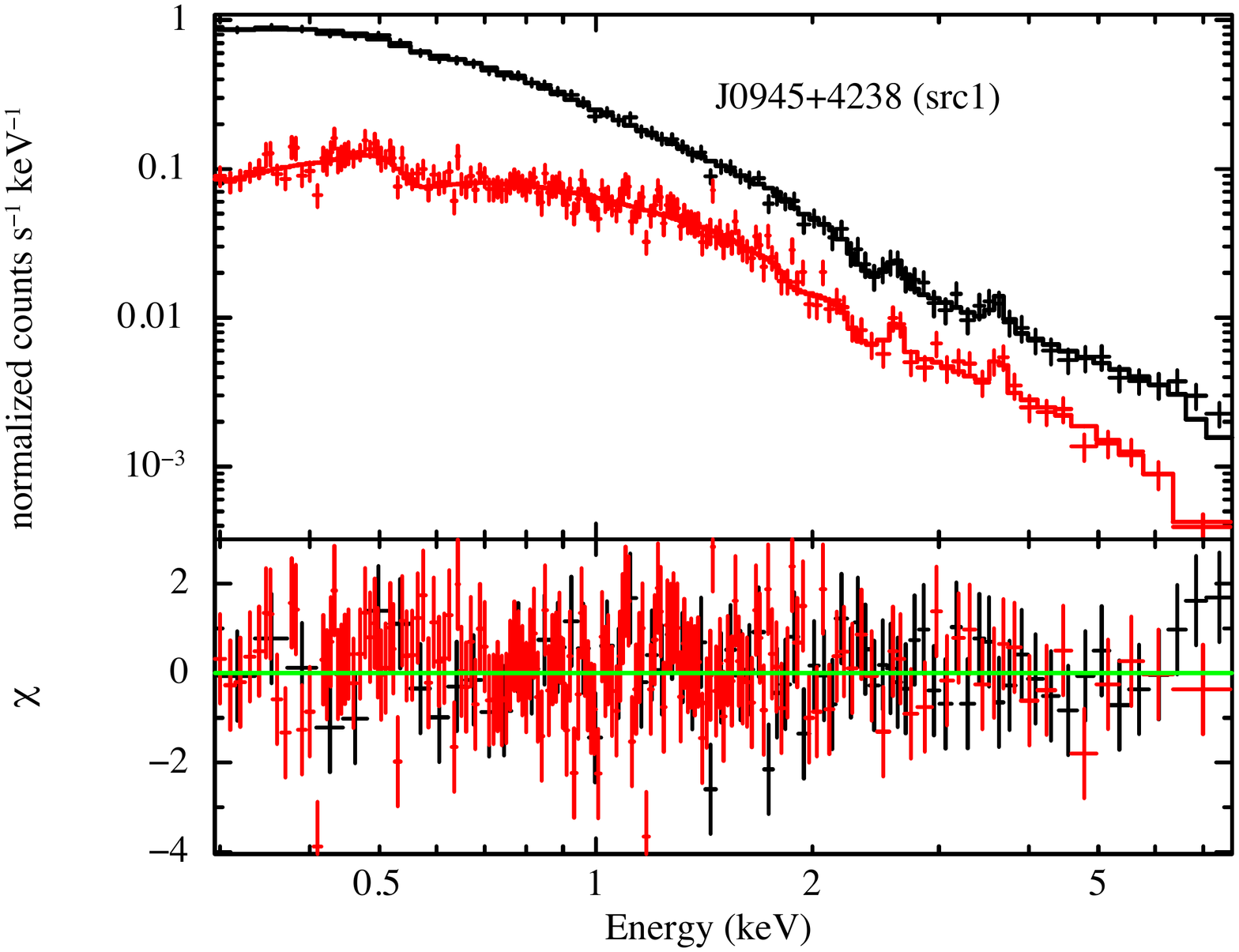}
\hspace*{-1.6cm}
\includegraphics[scale=.31,angle=-90]{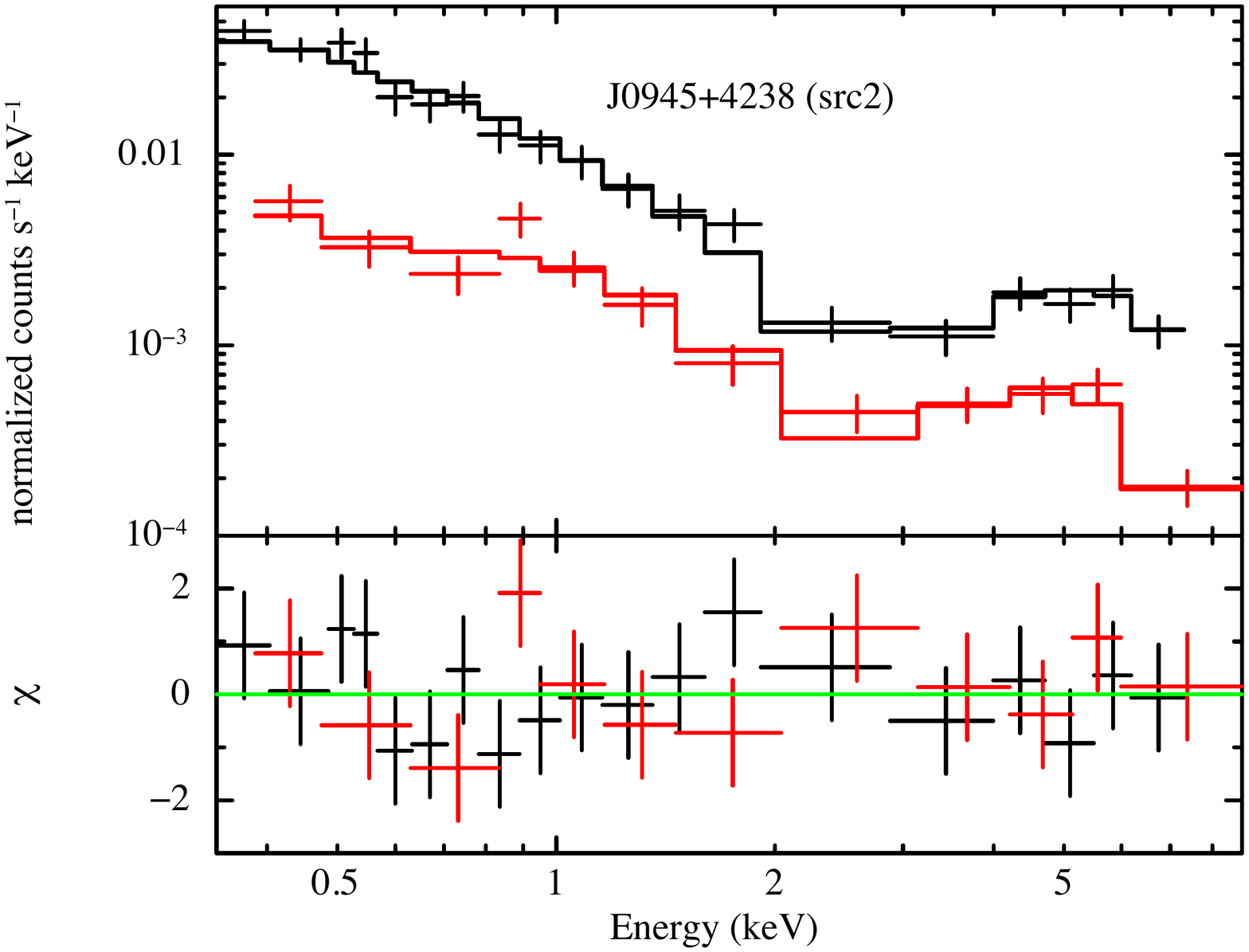}
\vglue-1cm
\includegraphics[scale=.31,angle=-90]{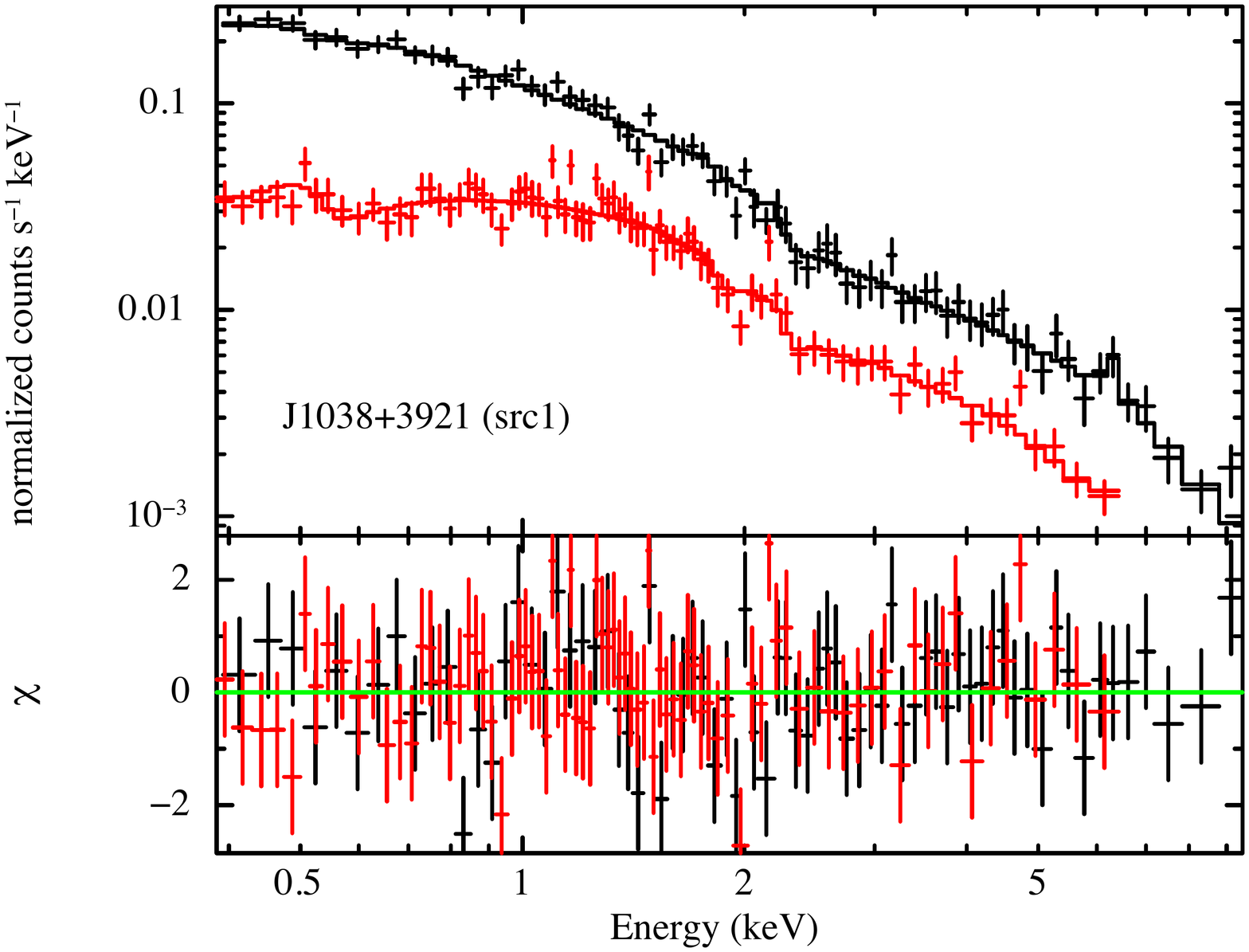}
\hspace*{-1.6cm}
\includegraphics[scale=.31,angle=-90]{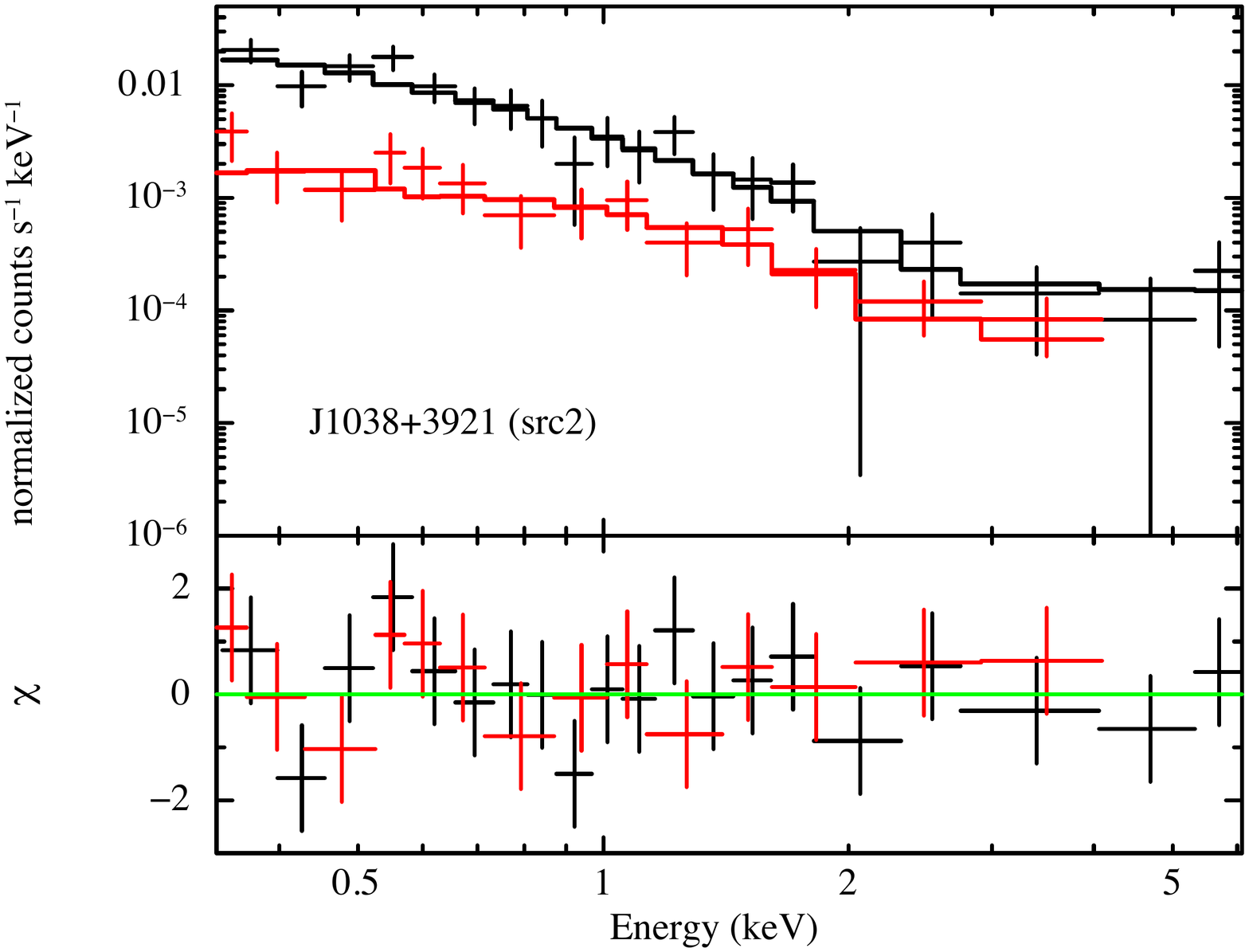}
\vglue-1cm
\includegraphics[scale=.31,angle=-90]{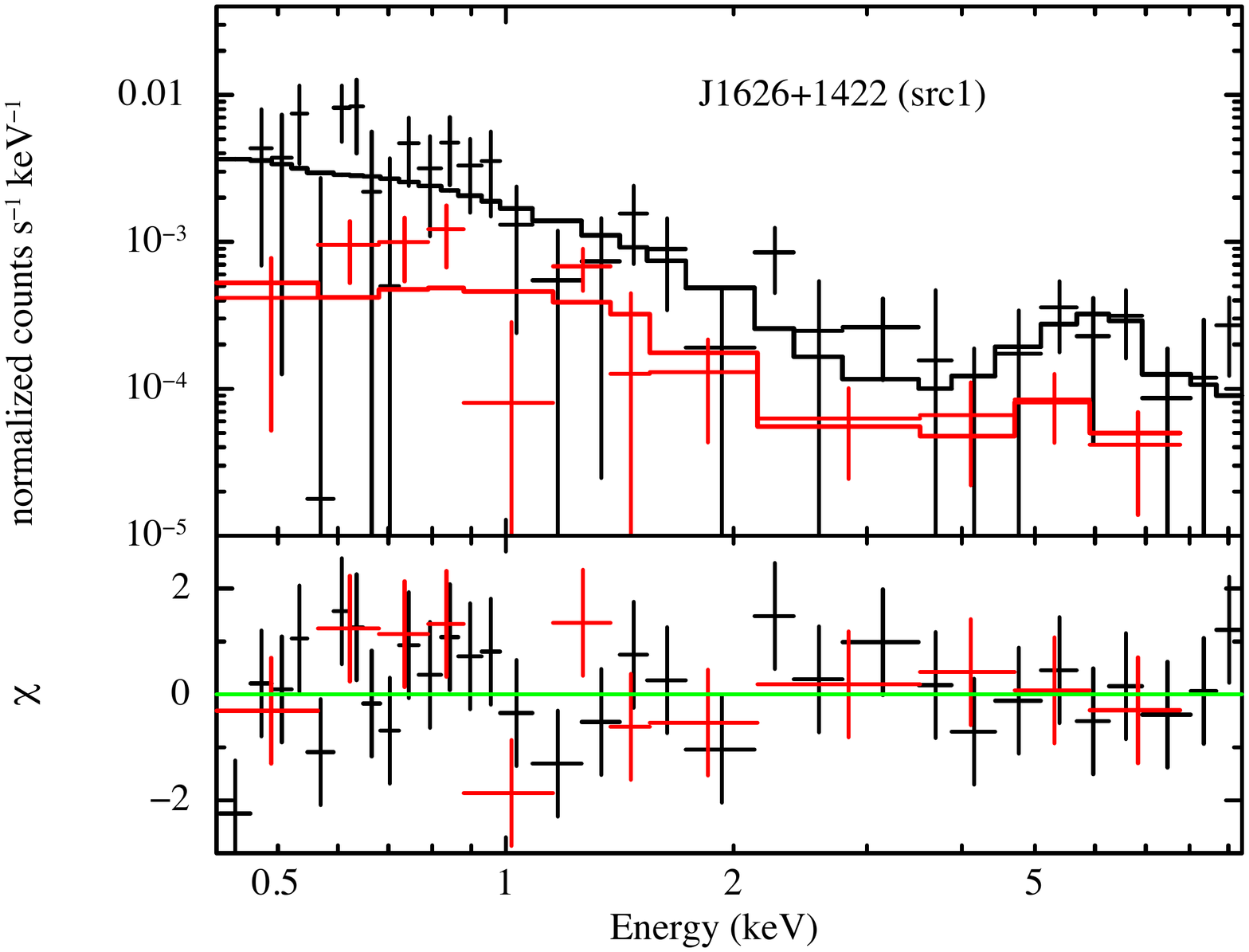}
\hspace*{-1.6cm}
\includegraphics[scale=.31,angle=-90]{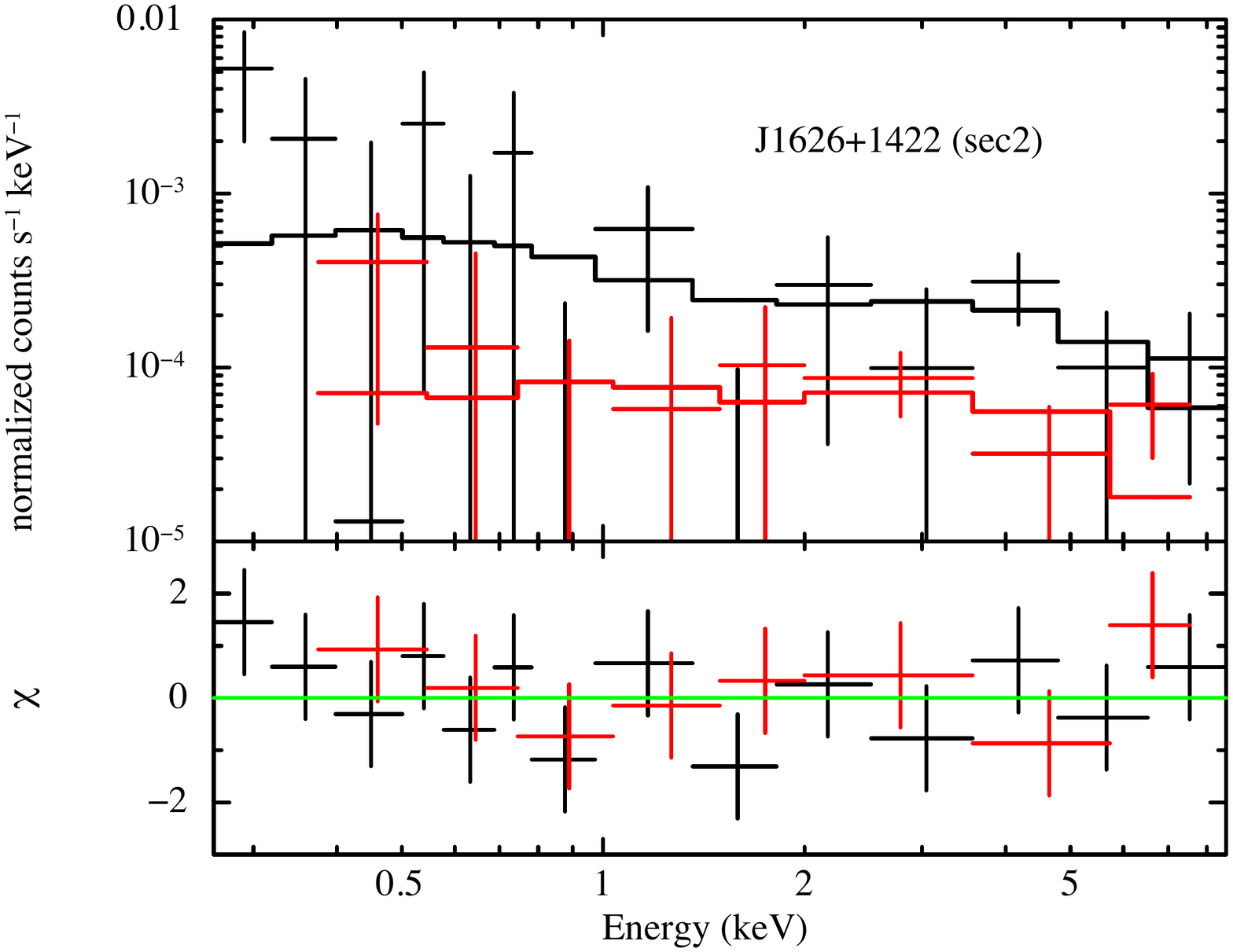}
\vglue-1cm
\includegraphics[scale=.31,angle=-90]{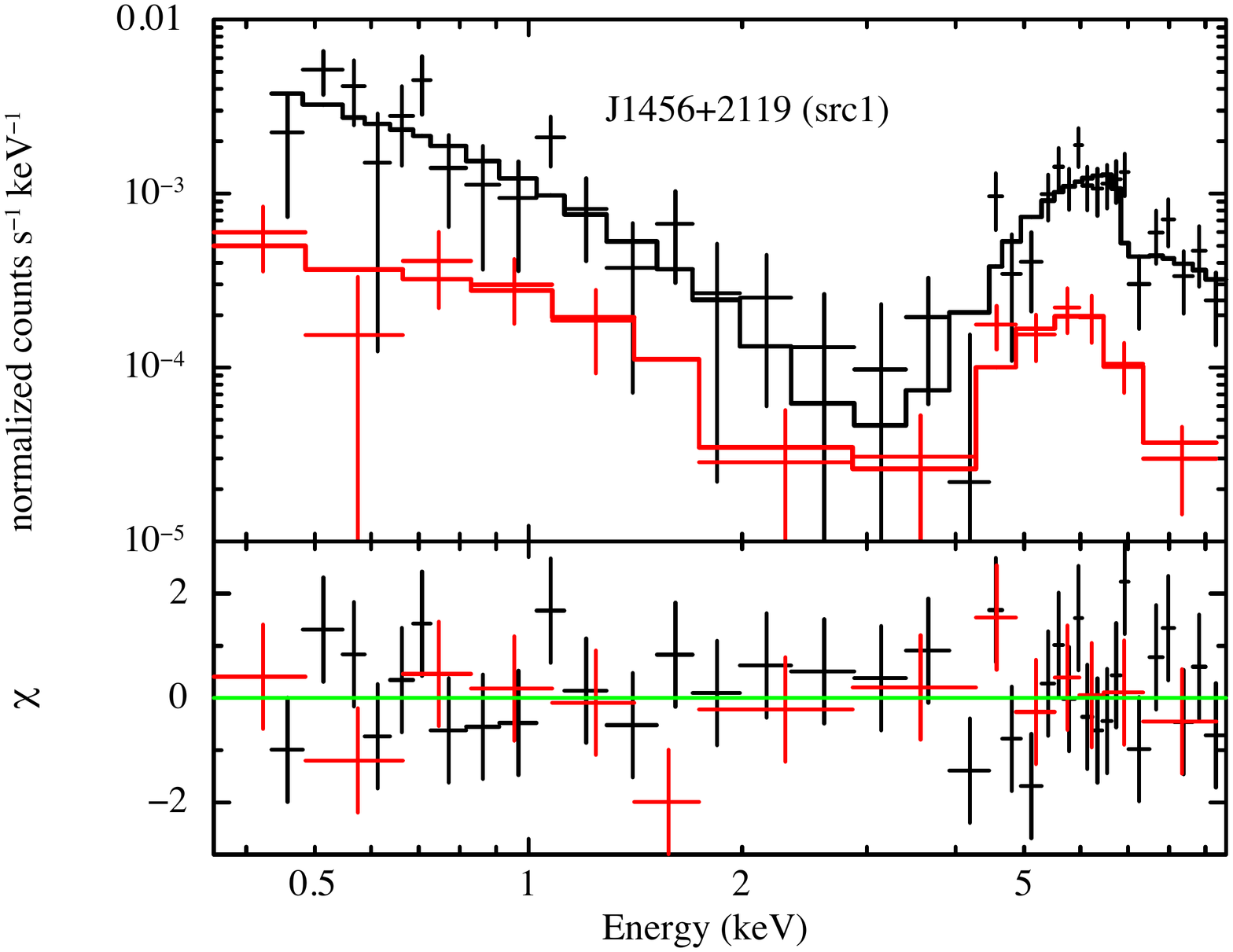}
\hspace*{-1.6cm}
\includegraphics[scale=.31,angle=-90]{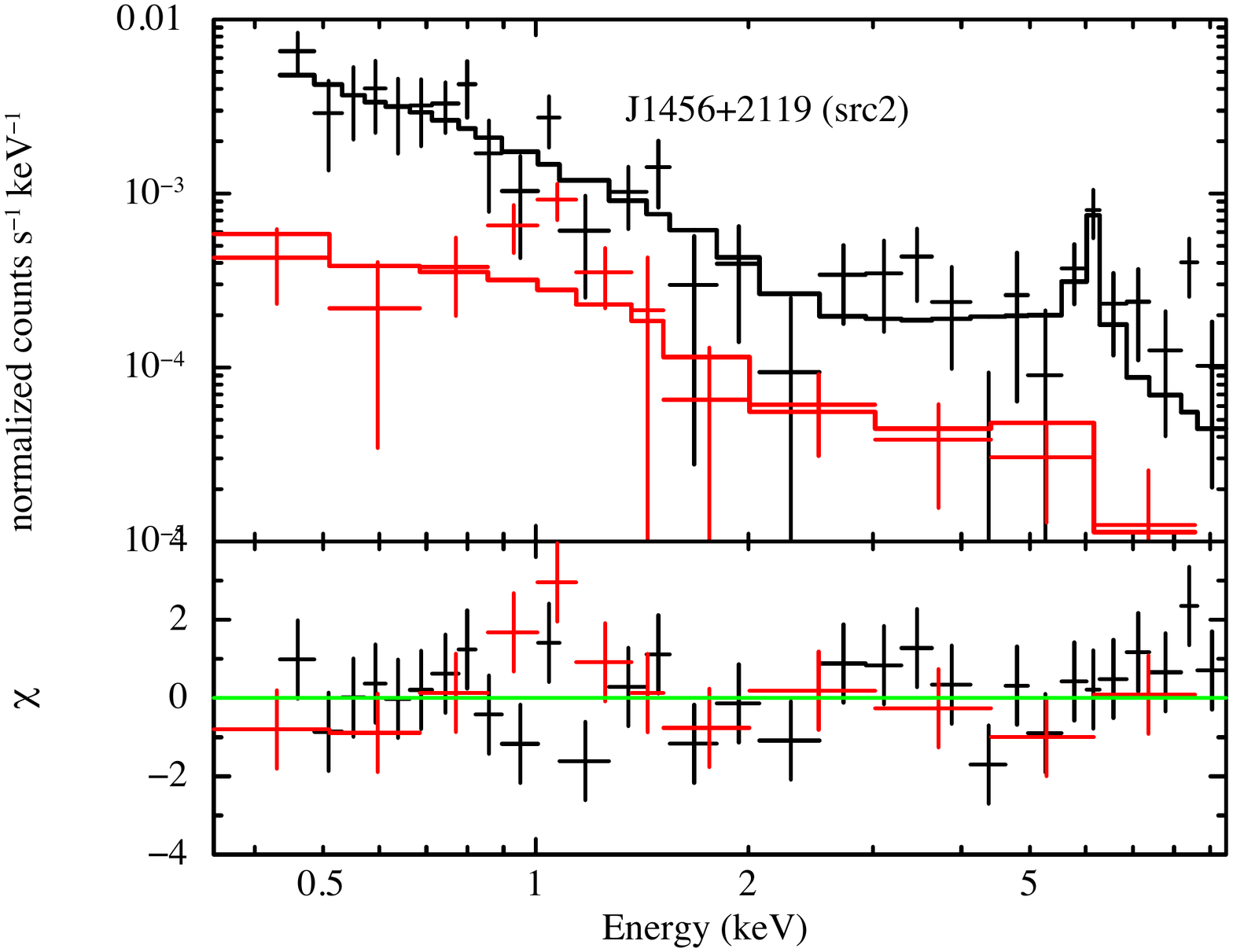}
\vglue-0.15cm
\caption{\xmm\ spectra of the eight sources. From top to bottom: SDSS J0945+4238, SDSS J1038+3921, SDSS J1626+1422, SDSS J1456+2119. Spectral data and best-fit models are reported in the top panels (EPIC-pn: black; MOS12: red), while data-to-model residuals (in units of $\sigma$) are shown in the lower panels.}
\label{fig:xmm_euf}
\end{figure*}

In the following we report the X-ray properties of each dual system and compare them with the classification obtained from the analysis of the optical spectra. In Fig.~\ref{fig:xmm_euf}, we plot the pn+MOS12 best-fitting spectra (upper panels) and residuals (lower panels) for all four systems. Best-fit values for all systems are reported in Table \ref{tab:xmm_analysis}.

\subsubsection{SDSS J0945+4238}
The sources in J0945+4238 are detected with a S/N ratio of 100 (src1) and 25 (src2) in the 0.3--10~keV band. The soft (0.3--2 keV) and hard (2--8 keV) band images (see Fig.~\ref{fig:xmm_reg}, upper left panels) are suggestive of the presence of heavy obscuration in src2. 
Given the relative proximity between the two sources (21\arcsec), we estimated that the PSF wings of src1 contaminates src2 in the hard band by $\le$20\%. 

When fitted with the BLM model, src1 does not require any absorption component (\nh $\le$ 3$\times$10$^{20}$\cm2, see Table~\ref{tab:xmm_analysis}) in addition to the Galactic one, and the X-ray spectrum looks like a type~1 Seyfert, in agreement with the optical classification (see Table~\ref{tab:sdss}). The residuals with respect to a power law model plus a thermal component still show an excess in the hardest X-ray band and in the Fe line region around 6~keV. The fit marginally improves ($\Delta\chi^2$=5) by adding a cold reflection component (\texttt{PEXRAV} model in {\sc xspec});  the best fit of the reflection fraction is R=2.2$\pm$1.8. 
The fit further improves by adding two narrow emission lines, whose best-fit energies are 2.81$\pm$0.03~keV and 3.91$\pm$0.07~keV, with equivalent width of EW=80$\pm$30~eV and 110$\pm$50~eV, respectively. These lines, clearly visible in both pn and MOS12 data, are statistically significant (the F-test probability is 99.7 per cent for both lines); furthermore, we note that these features lie in an energy range where no calibration issues should be present (i.e., the effective area of pn and MOS cameras is smooth).  
The closest transitions at these energies are are Si XV or Si XVI (2.45 keV and 2.62 keV) and Ca XVIII or Ca XIX (3.9 and 4.1 keV), nevertheless we cannot provide a physically convincing explanation for their presence of both lines in the J0945+4238 src1 spectrum. On the basis of the available optical imaging data, it seems unplausible that these emission lines come from other sources in the X-ray source extraction region of src1.

The X-ray spectrum of src2 is typical of a type~2 AGN, as expected by the optical diagnostics (see Fig.~\ref{fig:bpt}), with an X-ray obscured primary continuum and a soft component. 
Fitting src2 with the BLM provides good results: the primary continuum is obscured by a column density of N$_{\rm H}=24^{+8}_{-6}\times10^{22}$ \cm2. 
The soft power law contributes more that 10 per cent to the emission below 2~keV (f$_{\rm scatt}\sim$ 10). However, this value must be considered just an upper limit due to the possible contamination of src1.

The system SDSS~J0945+4238 therefore comprises a NLSy1, X-ray unobscured AGN, and a type~2 Compton-thin AGN. 

\subsubsection{SDSS J1038+3921} 
The nuclei in J1038+3921 are detected in 0.3--10 keV energy band with a S/N of 68 and 13, respectively (see Fig.~\ref{fig:xmm_reg}, upper rigt panels).

When fitting src1 data with the BLM model we only find an upper limit to the absorption component (N$_H\le3\times10^{20}$~\cm2), and the X-ray spectrum is consistent with expectations in case of a type~1 Seyfert galaxy. This is also in agreement with the optical classification (see Table~\ref{tab:sdss}).
A narrow Fe~K$\alpha$ line is marginally (2$\sigma$) required by the fit; its rest-frame energy is 6.6$^{+0.2}_{-0.3}$~keV and the equivalent width is EW=130$^{+150}_{-110}$~eV. 
Including a cold reflection component provides a negligible improvement ($\Delta\chi^2$=2) in the spectral fit, with the best-fit reflection fraction R=2.0$^{+2.0}_{-1.5}$, and the photon index being still consistent with the previous value within errors. 

Fitting src2  with the BLM, we obtain a very flat primary photon index above 
2~keV, $\Gamma\sim-1$, and a measured 2--10 keV luminosity of $\sim3\times10^{40}$~\ergs. The Balmer decrement (H$\alpha$/H$\beta \sim$ 3.9) leads to a de-reddened [OIII] luminosity of about 2 $\times$10$^{41}$~\ergs\ (see Table~\ref{tab:sdss}) which, in turn, gives an X-ray to optical luminosity ratio of $\log(\rm {L_x/L_{[OIII]}})=-0.9$. As described above, this low value is strongly indicative of a heavily obscured, possibly Compton-thick AGN (e.g., \citealt{lamastraetal09,vignalietal10}). We test then the possibility that the direct X-ray radiation is totally absorbed by a Compton-thick medium (N$_{\rm H}\ge10^{24}$~\cm2) and that we are observing is the reflected component in the \xmm\ observing band.  
To this goal, we replace the absorbed hard X-ray primary continuum with a pure reflection spectrum (\texttt{PEXRAV} in {\sc xspec}), as expected in case of a Compton-thick AGN, and leave a soft X-ray scattered component dominating at low energies. This model typically provides a good representation of the X-ray spectra of highly obscured AGN, at the level of statistical quality of the objects discussed in this paper (e.g., \citealt{lanzuisietal15}).
The fit obtained using this model is good: the observed luminosity is $\sim10^{41}$~\ergs\ and no significant residuals are visible (see Fig.~\ref{fig:xmm_euf}).
Assuming the intrinsic X-ray luminosity to be 80 times the observed L$_{\rm x}$ \citep{lamastraetal09,marinuccietal12}, we obtained L$_{\rm x}$=8$\times$10$^{42}$ \ergs\ and $\log(\rm{L_X/L_{[O\ III]}})=1$, in a very good agreement with the values found in type~2 Seyfert galaxies once the observed luminosities are corrected for the obscuration. 		
The observed spectrum below 4~keV is completely dominated by the soft power law component. However, when considering the estimated intrinsic luminosity, the scattered component f$_{\rm scatt}$ contributes about 1\% to the 0.5--2 keV luminosity, but this should be considered as a lower limit, due to the fact that the primary continuum below 10 keV is completely obscured from the compton thick gas. 

Summarizing, the system SDSS~J1038+3921 consists of an optical type~1, X-ray unobscured AGN, and one Compton-thick AGN optically associated with a type~2 AGN. 

\subsubsection{SDSS J1626+1422} 
Both sources in the SDSS~J1626+1422 system are detected in the broad band with a signal-to-noise ratio of 15 and 10 (see Fig.~\ref{fig:xmm_reg}, bottom left panels), respectively.
For what concerns src1, the low signal-to-noise ratio of the data does not allow us to constrain the primary photon index, which we fixed to the typical value found in Seyfert galaxies ($\Gamma=1.9$ e.g. \citealt{bianchietal09}). When the BLM is applied to its EPIC spectra, src1 appears highly obscured, with \nh\ in the range $\sim(20-220)\times10^{22}$~\cm2.
Correcting the observed X-ray luminosity for the obscuration (best-fit value of \nh=67$\times$10$^{22}$~\cm2) we obtain 8 (6)$\times$10$^{41}$~\ergs\ in the 2--10 (0.5--2) keV band, with the soft component contributing $\sim3$\% to the 0.5--2 keV unabsorbed luminosity.
As in the case of SDSS J1038+3921 (src2), we fit the data replacing the transmission model with a pure reflection model in the 2--10 keV band, while the soft X-ray data still modelled with a power law. The fit is still good ($\chi^2$/dof=38/39) and the observed 2--10~keV luminosity is about 10$^{41}$~\ergs, which translates into an intrinsic luminosity of 8$\times$10$^{42}$~\ergs\ by adopting the factor 80 (valid for heavily obscured sources) as described above. 
In this case the soft X-ray component would contribute to $\sim$3\% (f$_{\rm scatt}$) of the total unabsorbed luminosity in the 0.5--2 keV band.
We cannot statistically distinguish among the two scenarios (Compton thick \textit{vs} -thin), however the solid conclusion is that the \xmm\ spectrum of src1 is consistent with a highly obscured, possibly Compton-thick AGN.

We fit src2 \xmm\ data with the BLM model; given the low quality of the spectrum, we tied the  soft photon index to the primary and fixed their value to 1.9 as measure in a larger sample of LINER galaxies \citep{gonzalezetal09}.
Although we only have an upper limit for src2 in the soft band (below 2 keV), the hard X-ray data allowed us to reproduce the shape of the continuum and measure the absorption column density.
The source is absorbed by cold gas, with \nh$\sim6\times10^{22}$~\cm2; the derived intrinsic, rest-frame 2--10~keV luminosity is 1.2$\times10^{41}$~\ergs, and 
$\log(\rm{L_x/L_{[OIII]}})=1.5$ is within the range expected for an AGN obscured by Compton-thin gas.
Optical diagnostic diagrams suggest this source to be a LINER; however, the derived X-ray luminosity, at face value, is above the mean value found in the systematic X-ray study of the largest sample of LINERs thus far carried out  \citep{gonzalezetal09}. In their investigation, \cite{gonzalezetal09} have shown that when LINERs contain an AGN their average 2--10~keV luminosity is $\log\ {\rm L_x}=40.22\pm1.24$, higher than in cases when the powering source is not clearly associated with an active nucleus ($\log\ {\rm L_x}=39.33\pm1.16$). 

Summarizing, SDSS~J1626+1422 system comprises one Compton-Thick AGN candidate (optically classified as a type~2 AGN) and an X-ray obscured LINER, likely accretion-driven on the basis of the relatively high X-ray luminosity. 

\subsubsection{SDSS J1456+2119} 
 The dual AGN in the SDSS~J1456+2119 system are detected with a signal-to-noise ratio of 20 and 18 (see Fig.~\ref{fig:xmm_reg}, bottom right panels). respectively.
Fitting src1 with the BLM provides a good fit: the source is highly obscured, with 
N$_{\rm H}=75^{+28}_{-23}\times10^{22}$~\cm2.  
The spectrum below 2~keV is reproduced using a soft unabsorbed power law with steep photon index ($\Gamma_{\rm soft}\sim2.6$). The contribution of this component to the 0.5--2~keV intrinsic luminosity is f$_{\rm scatt} \sim$ 0.7\%. 
It is worth noting that a third X-ray source is visible close to src1 ($\sim$23~arcsec separation); we extracted and analised the X-ray spectrum of this source.
Due to the \xmm\ EPIC PSF ($\sim$90\% of the encircled energy fraction at 1.5 keV lies in a circular region of 30\arcsec\ radius), this source only marginally ($<$25\%) contaminates the 0.5--2~keV spectrum of src1. 

Fitting src2 with the BLM provides an unphysically hard photon index ($\Gamma=-1$) and 
an X-ray luminosity of $\sim$10$^{41}$~\ergs. The presence of heavy obscuration is confirmed by the X-ray vs. de-reddened [OIII] (see Table~\ref{tab:sdss}) luminosity ratio, $\log(\rm{L_X/L_{[OIII]}})=-0.5$. 
Fitting the \xmm\ data with a reflection-dominated model, we obtain a good fit and no systematic residuals are observed (see Fig.~\ref{fig:xmm_euf}). 
The observed 2--10~keV luminosity of 10$^{41}$ \ergs\ would translate into an intrinsic luminosity of 8$\times10^{42}$~\ergs\ (using the factor 80 discussed previously), and a $\log(\rm{L_X/L_{[OIII]}})=1.3$. 
A narrow Fe~${\rm K\alpha}$ line is marginally (3$\sigma$) required by the fit. Its rest-frame energy is 6.3$^{+0.2}_{-0.1}$~keV, and its EW=0.98$^{+0.86}_{-0.43}$~keV is consistent with the expectations in case of heavily obscured objects.  
The soft X-ray spectrum is well modelled with a steep power law, whose contribution to the 0.5--2 keV intrinsic luminosity is f$_{\rm scatt}\sim$3\%. 

\xmm\ data of system SDSS~J1456+2119 are therefore consistent with a pairs of obscured AGN, one in the Compton-thin and one in the Compton-thick regime. 

\section{Discussion} 
\label{sec:disc}

In this work we characterised the X-ray properties of an optically selected sample of dual AGN systems, drawn from SDSS \citep{liuetal11} and classified using standard emission-line ratio diagnostics \citep{baldwinetal81}. 
With the main goal of characterising their multi-wavelength properties, we started an observational campaign of a well defined sample of dual AGN with projected separation below 60 kpc.
In this paper we reported on the study of four systems at 
kpc-scale separation observed with SDSS and \xmm.
Because \cite{liuetal11} used a standard automatic procedure to classify sources in their SDSS sample, we carefully re-analysed the optical spectra, and the results of our fitting procedure are reported in Sect. \ref{sec:data_analysis}.

\subsection{The emission properties of dual AGN sample} 
Through X-ray data we confirm the presence of two AGN in all analysed systems, and the X-ray classification is in perfect agreement with the optical classification, as predicted by the AGN unification scenarios \citep{antonucci93} X-ray unabsorbed AGN are associated with broad-line AGN (BL AGN), and X-ray obscured sources with narrow-line AGN (NL AGN) in the optical band. 
The X-ray spectra are reproduced with an absorbed primary power law and a second unabsorbed component in the soft X-ray band ($<$ 2 keV) to reproduce the soft excess.
Additional spectral components to model the reprocessing of the nuclear radiation (e.g., Fe emission line and Compton reflection) are also introduced to model the two type~1 Seyfert (J1038-src1 and J0945-src1).
The range of bolometric luminosity in our sample is log(L{$_{\rm bol}$/\ergs)}$\sim$42--44.1, and the average value of the primary photon index and intrinsic X-ray luminosity (in the 2--10 keV band) measured in our sample is 1.91 (when left free to vary in the fit, with 0.19 standard deviation) and 3.7 $\times 10^{42}$~erg~s$^{-1}$ (0.19 standard deviation), respectively. These values are in a very good agreement with those measured in larger samples of AGN in isolated systems         
\citep{bianchietal09,derosaetal12}.

The soft component (soft excess) in the type~2 AGN is reproduced with a scattered power-law and contributes about f$_{\rm scatt}\sim$ 0.5--10\% to the total intrinsic luminosity in the 0.5--2~keV band in our sample. 
This value is in good agreement with the fraction of soft X-ray emission component commonly observed in type~2 AGN, and is likely due to the emission from the NLR  \citep{bianchietal06}. However, it could be also produced through Thomson scattering of the primary X-ray emission in photoionized gas \citep{uedaetal07}. Finally, soft X-ray emission could also be due to star-forming regions from a population of X-ray binaries \citep{ranallietal03}. 
Low photon statistics in the soft band does not allow to disentangle individual contributions from star formation and photoionized gas. To estimate the emission coming from star-formation activity, we used the radio flux density at 1.4~GHz from the FIRST survey (see Table \ref{tab:obs}) as an indicator for SFR \citep{condon1992}. In four over eight sources we obtained an upper limit for the SFR, while for four sources an estimate of the SFR was derived. Then we used the SFR \textit{vs} L$_{\rm x}$ relation of \cite{ranallietal03} to evaluate the contribution of the SFR to the X-ray luminosity. The contribution of the SFR to the X-ray emission ranges from 1 to 10 per cent in our obscured sources, then we conclude that the SFR has a limited effect on the X-ray luminosity.

The origin of the soft excess in type~1 AGN is still unclear and commonly interpreted as due to relativistic reflection from the innermost region in the accretion disc \citep{crummyetal06} or the effect of thermal Comptonization of the disc photons in a warm, optically thick corona \citep{magdziarzetal98,petruccietal18}.
We reproduced the soft excess in the unabsorbed AGN with a simple phenomenological model, i.e. thermal component (a blackbody). Although the black body is not a physical representation of the spectral feature,  it is able to reproduce the soft excess with respect to the power-law in both the two type 1 AGN.
We found a temperature kT$\sim$0.1 keV for both 
type~1 AGN (J0945-src1 and J1038-src1); these are typical values found in Seyfert 1s \citep{gierlinskidone04}.
The fraction of the soft X-ray component with respect to the primary emission in the 0.5--2 keV band for the two type~1 AGN is $\sim$30 and $\sim$70 per cent, respectively. 
From our optical analysis (see Table~\ref{tab:sdss}) we classified J0945+4238 src1 as a NLSy1, which is expected to have a larger soft X-ray component with respect to 
broad-line AGN, possibly due to a higher accretion rate and lower BH mass  \citep{boelleretal96,jinetal12}.
We used H$\alpha$ luminosity and FWHM (see Table \ref{tab:type1}) in order to compute the BH mass for the NLSy1 J0945-src1 obtaining 4$\times$10$^{6}$M$\odot$ and, in turn, an Eddington luminosity of 5$\times$10$^{44}$ \ergs\ and an Eddington ratio $\lambda_{\rm Edd}\sim$0.23 (assuming the bolometric correction given by \citealt{marconietal04}, see Table~\ref{tab:xmm_analysis}). This value is higher with respect to average values of $\lambda_{\rm Edd}$ measured in large sample of BL AGN in the same luminosity bin (mean $\lambda_{\rm Edd}$=0.033 with 0.38 dispersion, see e.g. \citealt{lussoetal12}). The Eddington ratio of J1038+3921 src1, $\lambda_{\rm Edd}\sim$0.01 (see Table \ref{tab:type1}), is perfectly consistent with larger sample of BL AGN.
The spectral index of the primary power-law in J0945+4238 src1 is significantly higher than the other sources, 2.4 $vs$ 1.9, as also observed in X-ray spectra of NLSy1 \cite[and references therein]{jinetal12}.

\subsection{The nature of obscuration} 
Six out of the eight sources (75 per cent of the sample) show evidence for high obscuration by cold gas (column density \nh \gtsima 10$^{23}$ \cm2). The number of AGN with \nh \gtsima 10$^{24}$ \cm2 is in the range 2--4 when 90 per cent confidence errors on the column density are considered (see Table~\ref{tab:xmm_analysis}).
Although our sample is definitely too small to make any strong conclusion in terms of statistical incidence of obscured AGN in dual systems, we note that the fraction of CT AGN in our sample of dual AGN is 25--50 per cent, i.e. higher than the fraction of hard X-ray selected CT AGN in isolated systems in a similar luminosity interval (BAT: 27$\pm$4 per cent,  \citealt{riccietal15}, \nustar: $\sim$4 per cent, \citealt{marchesietal18}.) 
Recently, it has been suggested \citep{riccietal17} that AGN in multiple systems in a late stage of merging (with a projected distance d $\leq$ 10~kpc) are often type-2 AGN due to the huge amount of gas and dust obscuration channelled towards the nuclear region by the close galaxy encounter. Furthermore, in a sample of 44 Ultra Luminous IR Galaxies (ULIRG) from the GOALS survey \citep{sandersetal03} in the local universe, it has been demonstrated that the fraction of CT AGN in late mergers is higher (65$^{+12}_{-13}$ per cent) than in local isolated hard X-ray selected AGN (27$^{+4}_{-4}$ per cent, \citealt{riccietal17}), while it is marginally higher than isolated AGN in early mergers (35$^{+13}_{-12}$ per cent). The increasing of the obscuration with the disturbed/interacting morphology of the galaxy host has been also observed up to higher redshift  (up to z=1.5) using X-ray \chandra\ data from deep surveys \citep{kocevskietal15}. 
These authors have analised 154 heavily obscured AGN and compared their morphologies with control samples composed of moderately obscured and unobscured AGN.
They  found that the  fraction of galaxies undergoing mergers is higher in AGN with N$_{\rm H}$ above 3$\times$10$^{23}$ \cm2 with respect to unobscured AGN with N$_{\rm H} < $10$^{22}$ \cm2 (7.8$^{+1.9}_{-1.3}$  $vs$ 21.5$^{+4.2}_{-3.3}$ per cent). 
X-rays \citep{kocevskietal15,riccietal17} as well as more recent mid-IR color selection \citep{satyapaletal17} strongly suggest that AGN in late state of merging ($<$10 kpc) are highly absorbed (but see also \citealt{ellisonetal17}), with \nh\ above 10$^{24}$ \cm2.
All these observational results suggest that CT AGN in dual systems correspond to a phase where the supermassive black hole is still accreting gas during an early stage of an interaction or merging.

The systems investigated here have projected separations larger than 10 kpc, in the range 30--60~kpc, so they might be pre-existing AGN-AGN pairs, not triggered by encounters.
They are likely undertaking a pre- or early-merging phase, where galaxies are separated (i.e., prior to the first encounter) and not yet strongly interacting. In this respect we also note that the system J0945+4238, containing the AGN pair at the closest separation (30 kpc, see Table \ref{tab:obs}) shows some evidence of interaction on galaxy scale, with the possible presence of a initial tidal bridge (see Fig. \ref{fig:SDSS_images} first panel on the left). 

\begin{figure*}
\includegraphics[scale=.40,angle=0]{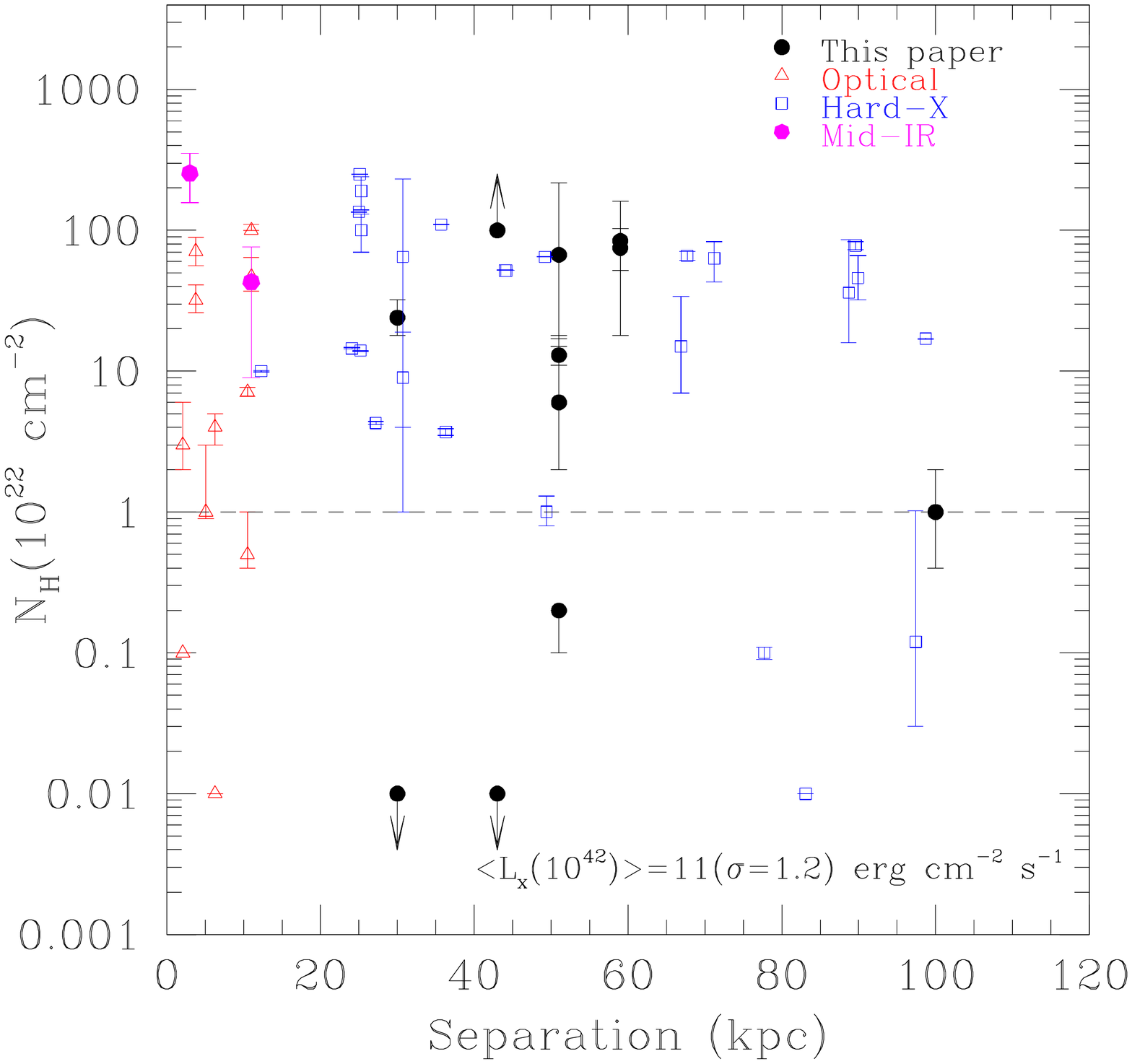}
\includegraphics[scale=.40,angle=0]{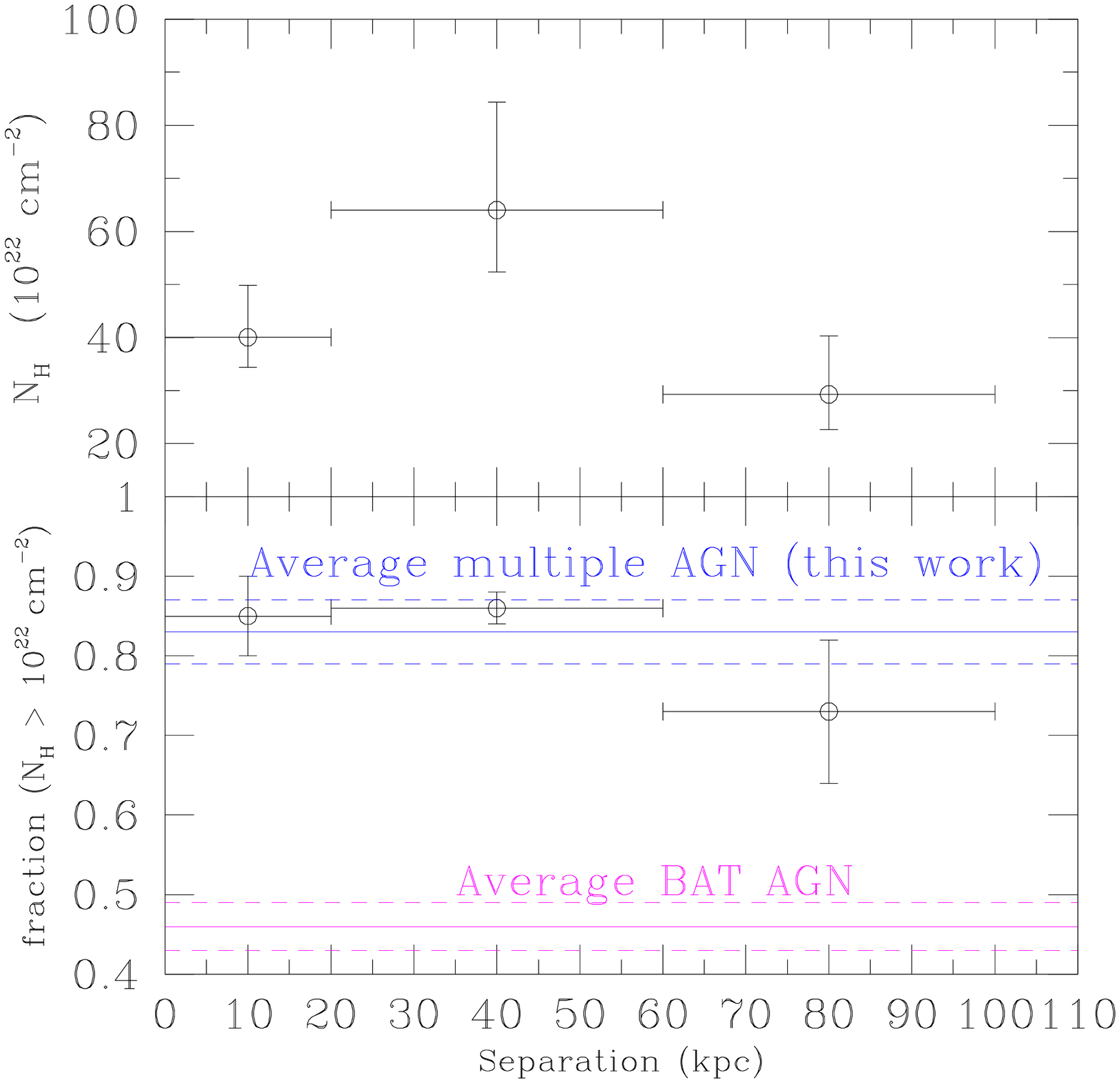}
\vglue-1.6cm
\caption{{\it Left}. Absorbing column density \nh versus projected separation between the nuclei for  known multiple systems from this work and literature. In particular, we used the systems investigated in this work and in \citep{derosaetal15} (black circles); from literature: (1) optically selected in SDSS \citep{liuetal13,bianchietal08,guainazzietal05,piconcellietal10} (red triangles); (2) hard X-ray selected mainly with \swift/BAT \citep{kossetal12} (blue squares); (3) IR selected  from GOALS \citep{riccietal17} (magenta circles).
{\it Right}. Upper panel represents the values of the absorption column density of the samples reported in the left diagram, averaged in 3 ranges of projected galaxies separations: 0--20, 20--60 and 60--100 kpc. Lower panel: Fraction of AGN in dual/multiple systems with \nh\ above 10$^{22}$ \cm2 as a function of the projected separation.  Blue lines represent the average value as obtained for the large sample investigated in this paper (61 sources, see left panels) when 90 per cent errors on \nh\ are taken into account. Magenta lines represent the \swift/BAT average values and 1$\sigma$ error as reported in \citet{riccietal15} in the same bin of 2--10~keV luminosity.}\label{fig:diagrams}
\end{figure*}

The small number of systems investigated here does not allow any strong statistical argument, therefore we decided to analyse additional sources selected in different wavebands and for which a measure of the absorption column density can be derived from X-ray observations.
In Fig. \ref{fig:diagrams} (left panel) we plot the absorption column density of AGN sources hosted in a dual/multiple system  as a function of the projected separation for the sample presented here and for additional systems observed in X-rays and taken from literature. We included the following dual AGN systems: (i) optically selected from the SDSS \citep{liuetal13,bianchietal08,guainazzietal05,piconcellietal10,derosaetal15}; (ii) hard X-ray selected as obtained by the \textit{Swift}-BAT survey \citep{kossetal12}; (iii) mid-infrared selected, as obtained by GOALS \citep{sandersetal03, riccietal17}. 
The total sample comprises 61 AGN with 2--10~keV average luminosity (11$\pm$2)$\times$10$^{42}$ \ergs\ (standard deviation $\sigma$=1.2$\times$10$^{42}$~erg~s$^{-1}$) and average N$_{\rm H}$=(50$^{+15}_{-9}$)$\times$10$^{22}$~\cm2 (standard deviation $\sigma$=4$\times$10$^{22}$~cm$^{-2}$).
In the right (upper) panel of Fig. \ref{fig:diagrams} we show the column density N$_{\rm H}$ in three different bins of projected separation, and in the lower panel the fraction of sources showing obscuring column density N$_{\rm H}>$10$^{22}$~\cm2. 

Although the available measurements of dual systems hosting AGN are sparse, it seems plausible that obscuration (\nh\ above 10$^{22}$~\cm2) is a common property also during the early stage of merging.
The fraction of AGN hosted in dual or multiple systems, up to large separation, with absorption larger than 10$^{22}$ \cm2 is 84$\pm$4 cent (see blue lines in the right-bottom panel of Fig. \ref{fig:diagrams}, accounting for the 90 per cent error on the N$_{\rm H}$ measurement). The fraction of isolated AGN that show absorption above 10$^{22}$ \cm2 as observed by BAT \citep{riccietal15} is 46$\pm$3 per cent (magenta lines Fig. \ref{fig:diagrams}, right-bottom panel). 
This comparison suggests that the ``environment" of dual AGN systems is different, namely more obscuring than that of isolated AGN. 
We do not find evidence for a trend of increasing N$_{\rm H}$ with decreasing separation (see the right-upper panel in Fig. \ref{fig:diagrams}).

\citet{blechaetal17} presented numerical simulations of AGN in galaxy mergers, showing an increase of \nh\ as the distance decreases. This effect is due to merger dynamics; the median value of \nh\ was about 3$\times$10$^{23}$\cm2, in good agreement with the value we have found in the large sample of dual systems investigated here (see Fig. \ref{fig:diagrams}).
However, their simulations only probe absorption on relatively large scales (above $\sim$ 50~pc, then absorption form the torus on pc-scale is not considered), therefore the \nh\ value should be considered as a lower limit.

Although  heavy absorption can prevent the detections of dual AGN (up to large scale separation) through X-ray surveys \citep{kossetal12}, our study demonstrates that a multiband approach provides a successful strategy to detect and characterise dual AGN in late and early stage of merging. 
In this scenario the type~1 unobscured Seyfert detected in J0945-src1 and J1038-src1 are unusual, especially considering their projected separation of few tens of kpc. Unobscured AGN have recently been detected with \chandra\ in a dual system at close separation ($\sim$8, kpc \citealt{ellisonetal17}).  
Studies of larger samples of X-ray selected AGN in galaxy pairs indicate a non-negligible fraction of X-ray unobscured AGN (Guainazzi et al., in preparation). It is intriguing to speculate that X-ray unobscured members of AGN pairs may represent systems where the efficiency of funnelling gas and dust toward the nuclear region is low, or where the nuclear environment has been cleared up by powerful AGN outflows. Our spectroscopic analysis unveiled a powerful outflow in a heavily obscured AGN (J1456+2119), that could constitute a step in a sequence of events, of which unobscured members of AGN pairs represent the terminal stage. 

Selecting AGN by their mid-infrared colors and emission-line properties, \cite{juneauetal13} investigated the star forming (SF) activity and morphology of obscured AGN. Galaxy mergers are expected to fuel AGN as well as intense episodes of star formation, with the AGN deeply obscured in a large amount of gas. In their analysis they found that the obscured AGN fraction is higher among galaxies with elevated SF rates, possibly due to galaxy interactions. This result is consistent with an evolutionary scenario in which the abundant gas supply in the obscured AGN phase enhances the star formation. The SF rate and the obscuration both decline as the fuel decreases \citep{alexander&hickox12,ellisonetal16}. 
However, a similar study is not possible in our four dual AGN systems because of the lack of far-IR data, needed to properly estimate the star-formation rates in their host galaxies. 

\section{Summary and conclusions}

In this work we have presented the optical and X-ray study on four dual AGN at kpc-scale separation.
They belong to a large sample of multiple AGN systems selected from SDSS \citep{liuetal11} and classified using standard emission-line ratio diagnostics. 
As a first step, our observational campaign characterises the properties of a well defined sample of dual AGN with projected separation in the range 30--60~kpc using SDSS and \xmm.  
In this paper we have focused on the X-ray absorption properties of the sample and how they compare to larger samples of dual AGN at various scale separations and to isolated systems.
Our main results are summarized below.

\begin{itemize}

\item Through standard optical emission-line ratio diagnostics (see Fig. \ref{fig:bpt}) all objects are identified as AGN except one that is a LINER; due to its relatively high X-ray luminosity, the LINER emission is likely accretion driven. 

\item All sources are detected in X-rays, either in the soft and/or high-energy band. 
The X-ray and optical classification perfectly matches each other: X-ray unabsorbed AGN are associated with broad-line AGN, and X-ray obscured sources with narrow-line AGN in the optical band. The X-ray spectral properties of all systems (L$_{\rm x}$ and $\Gamma$, when left free to vary in the fit) are in agreement with values found in large samples of isolated local AGN (see Table \ref{tab:xmm_analysis}).

\item Six over eight objects are obscured by cold gas ($\log$\nh > 23), and the fraction of Compton thick AGN ($\log$\nh > 24) is 25--50 per cent  (accounting for the 90 per cent error on the N$_{\rm H}$ measure). This fraction is higher than that found in isolated hard X-ray selected AGN (BAT: 27$\pm$4 per cent, \citealt{riccietal15}; \nustar: $\sim$4 per cent, \citealt{marchesietal18}).

\item In order to statistically probe systems at different  separations, we compared the absorption properties in our pairs with those  in larger samples observed in X-rays and selected in different ways (optical, IR and hard X-rays, see Fig. \ref{fig:diagrams}), for a total 61 sources. We find that the fraction of obscured (N$_{\rm H} \geq$ 10$^{22}$~\cm2) AGN  is 84$\pm$4 per cent up to large-scale separations ($\sim$ 100~kpc), i.e. higher with respect to isolated AGN as observed by BAT \citep[46$\pm$3 per cent]{riccietal15}.

\item  In this scenario, two systems presented here manifest a different behaviour, being composed of an unobscured type 1 AGN and an obscured (Compton-thick in one case) AGN. Although the number statistics is still too small, we can speculate that X-ray unobscured AGN in pairs may represent systems characterised by a low efficiency of funnelling gas and dust toward the nuclear region, or where the nuclear environment has been cleared up by powerful AGN outflows.

\end{itemize}

Although the sample we presented in this work is limited due to the paucity of measurements of dual/multiple AGN, our analysis is in agreement with the hypothesis that galaxy encounters are effective in driving gas inflow, thus increasing the obscuration \citep{kocevskietal15,riccietal17}.
In addition, our findings suggest that the environment of dual AGN systems is different - more obscured - with respect to isolated AGN up to wide separations ($\sim$ 100 kpc). 
We can speculate that these systems with larger amounts of obscuring gas have a high efficiency of funnelling gas and dust toward the nuclear region. Heavy absorption can prevent the X-ray detections of dual AGN and it has been proposed to explain  the disagreement between optical and X-ray fraction of mergers \citep{kossetal12}. 
Our analysis shows instead that the combined optical and X-ray observing strategy represents a powerful tool to select and confirm dual AGN systems, even if heavily obscured, in interacting galaxies.
A multiband approach, as the one proposed in the MAGNA project\footnote{http://www.issibern.ch/teams/agnactivity/Home.html}, is a successful strategy in order to detect and characterise dual AGN in late and early stage of merging.
Recently, we have obtained deep JVLA dual band radio and LBT data for the master sample galaxy pairs described in Sect. \ref{sec:sample} (in particular, LBT data for the systems at closer separation, below 10~arcsec). 
These data are being analysed with the aim of characterising their radio morphology and spectral behaviour, and will be presented in forthcoming papers (Herrero-Illana et al. in preparation, Husemann et al. in preparation).

\section*{Acknowledgements}
The authors would like to thank the referee for her/his constructive comments and suggestions. 
The authors, members of the MAGNA project (\texttt{http://www.issibern.ch/teams/agnactivity/Home.html}), gratefully acknowledge support of the International Space Science Institute (ISSI), Bern, Switzerland, and the hospitality of the Lorentz Center for international workshops, where one portion of this work was carried out. The Italian authors acknowledge financial contribution from the agreement ASI-INAF n.2017-14-H.O. CV acknowledges financial support from the Italian Space Agency under the contract ASI-INAF I/037/12/0. We thank the hospitality of ESA-ESTEC where part of this work has been carried out, 
BH is grateful for financial support from the ESA-ESTEC visitor program. \\
Funding for SDSS-III has been provided by the Alfred P. Sloan Foundation, the Participating Institutions, the National Science Foundation, and the U.S. Department of Energy Office of Science. The SDSS-III web site is http://www.sdss3.org/.
SDSS-III is managed by the Astrophysical Research Consortium for the Participating Institutions of the SDSS-III Collaboration including the University of Arizona, the Brazilian Participation Group, Brookhaven National Laboratory, Carnegie Mellon University, University of Florida, the French Participation Group, the German Participation Group, Harvard University, the Instituto de Astrofisica de Canarias, the Michigan State/Notre Dame/JINA Participation Group, Johns Hopkins University, Lawrence Berkeley National Laboratory, Max Planck Institute for Astrophysics, Max Planck Institute for Extraterrestrial Physics, New Mexico State University, New York University, Ohio State University, Pennsylvania State University, University of Portsmouth, Princeton University, the Spanish Participation Group, University of Tokyo, University of Utah, Vanderbilt University, University of Virginia, University of Washington, and Yale University. 




\bibliographystyle{mnras}
\bibliography{magna_library} 




\appendix
\section{Figures of the optical spectral modelling}
\label{sec:appendix}
\begin{figure*}
 \includegraphics[width=\textwidth]{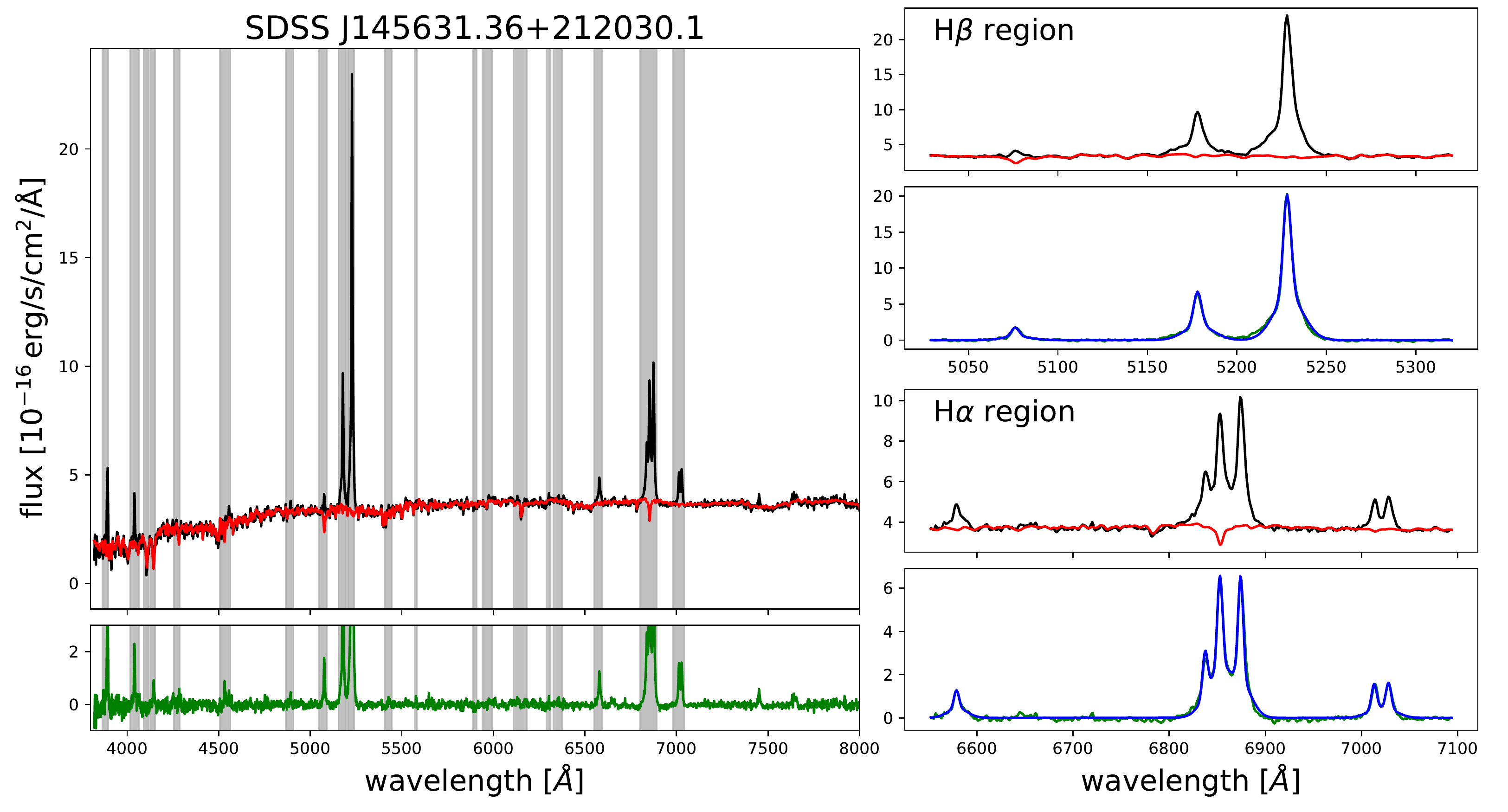}
 \caption{Same as Fig.~\ref{fig:type2_fitting} for SDSS J145631.36+212030.1.}
\end{figure*}
\begin{figure*}
 \includegraphics[width=\textwidth]{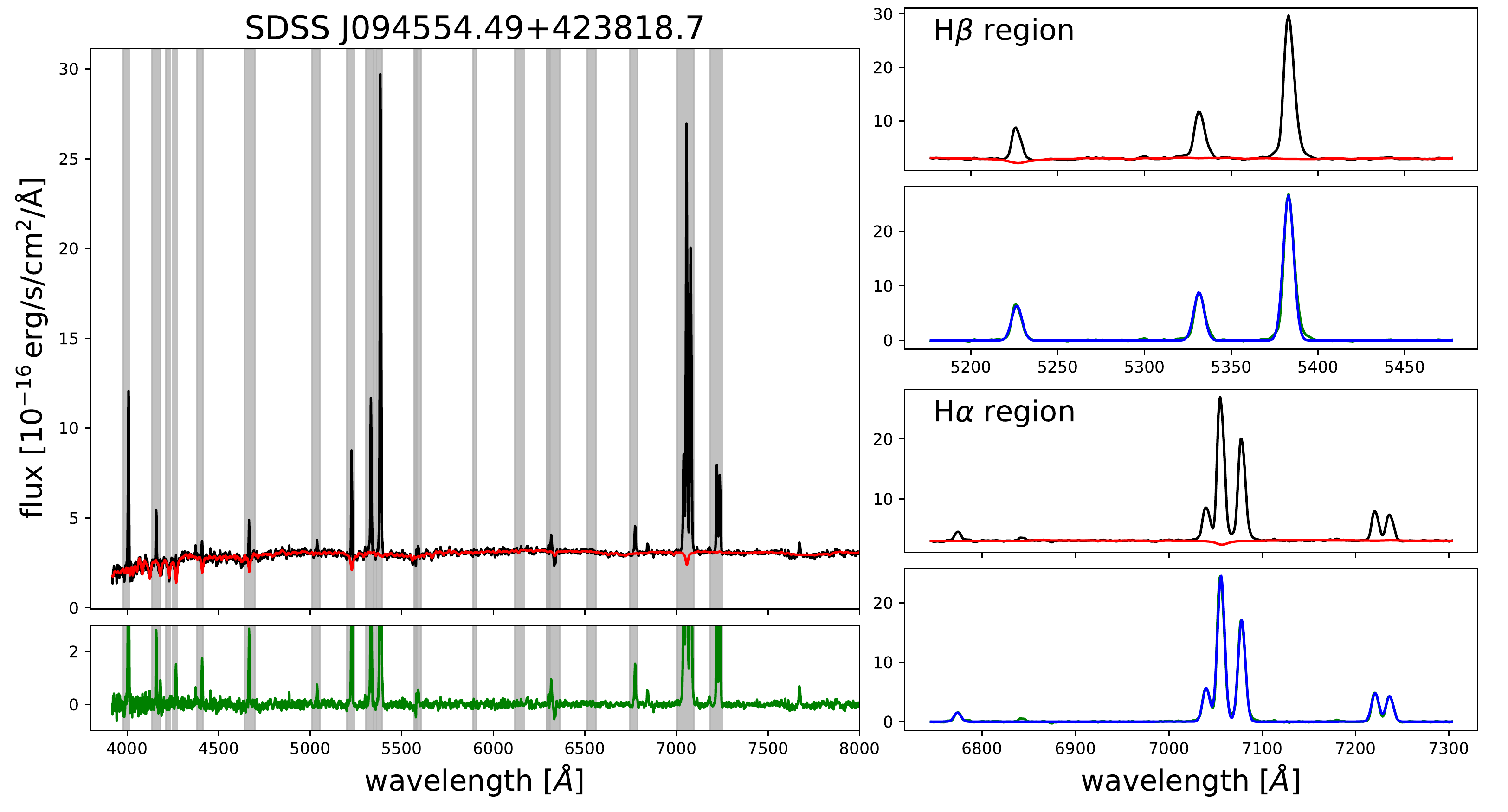}
 \caption{Same as Fig.~\ref{fig:type2_fitting} for SDSS J094554.49+423818.7.}
\end{figure*}
\begin{figure*}
 \includegraphics[width=\textwidth]{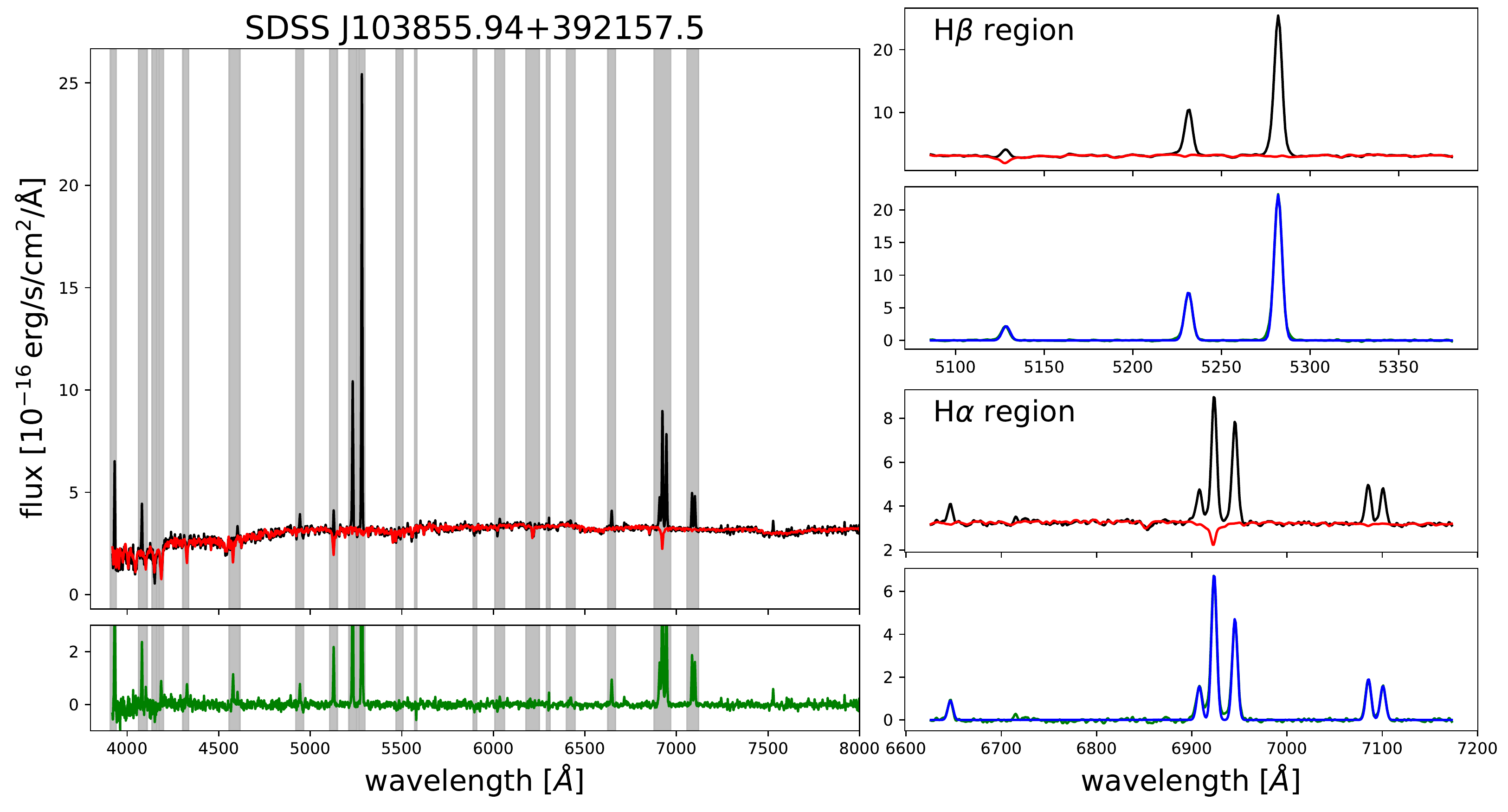}
 \caption{Same as Fig.~\ref{fig:type2_fitting} for SDSS J103855.94+392157.5.}
\end{figure*}
\begin{figure*}
 \includegraphics[width=\textwidth]{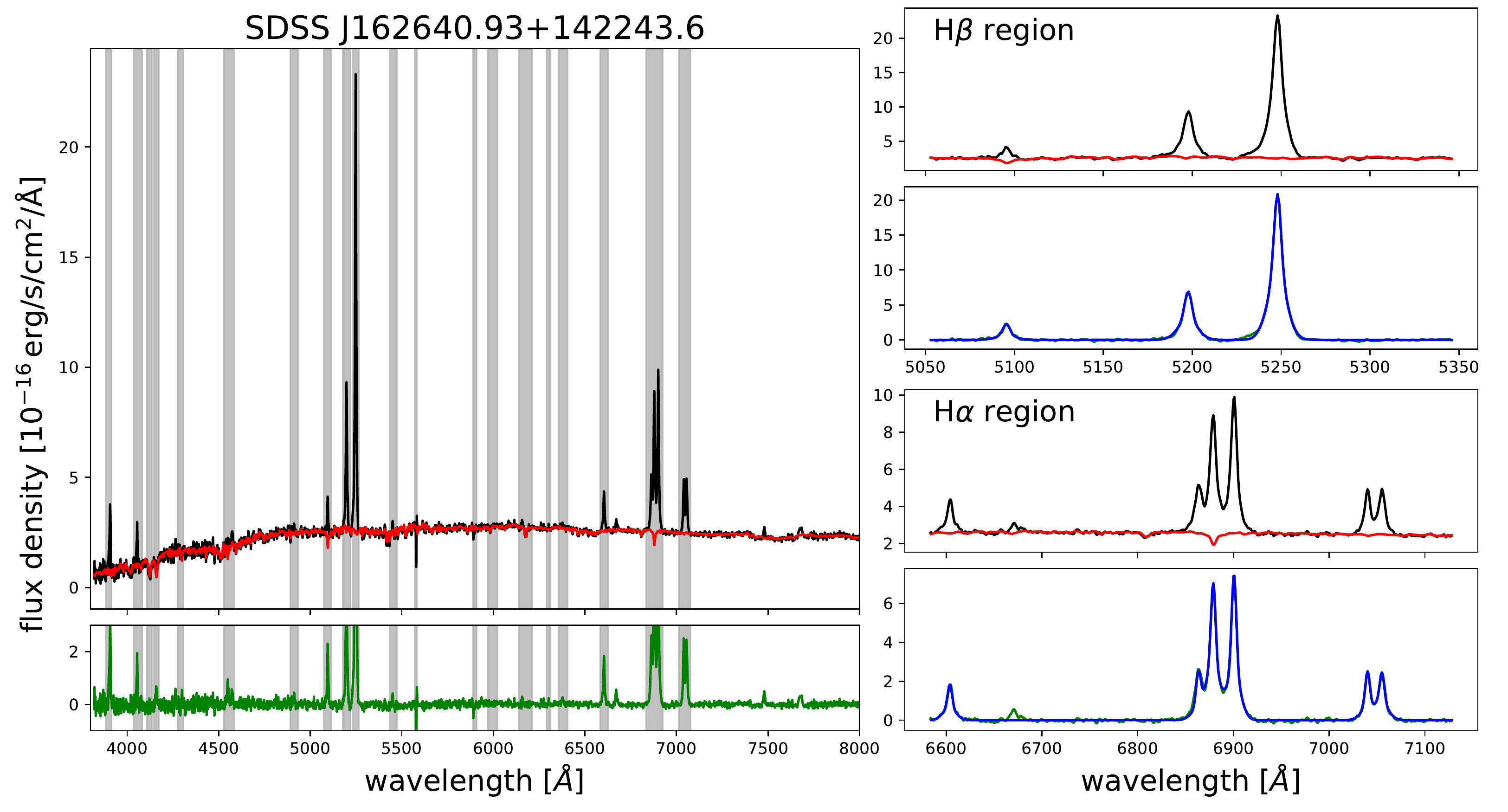}
 \caption{Same as Fig.~\ref{fig:type2_fitting} for SDSS J162640.93+142243.6.}
\end{figure*}
\begin{figure*}
 \includegraphics[width=\textwidth]{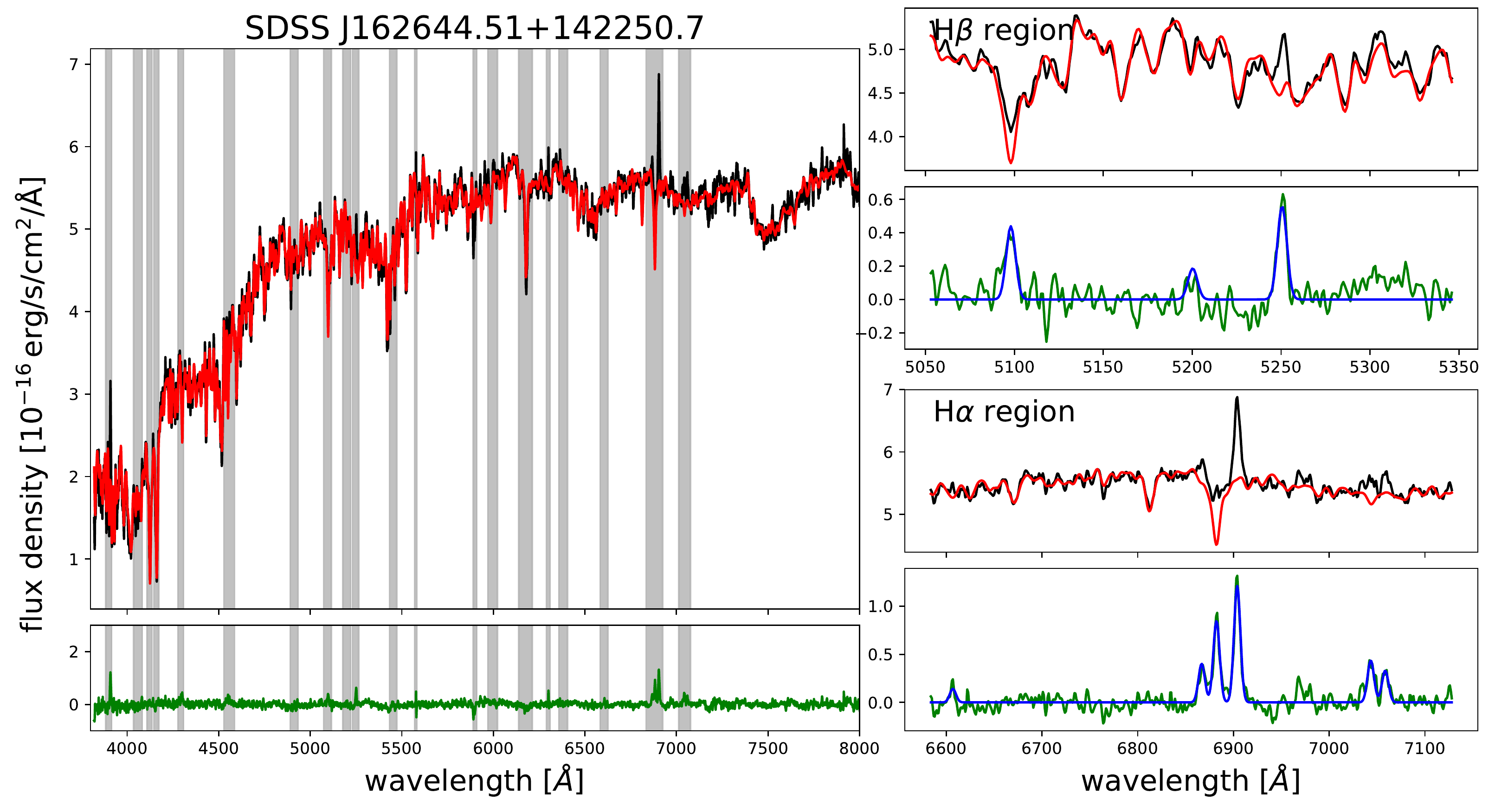}
 \caption{Same as Fig.~\ref{fig:type2_fitting} for SDSS J162644.51+142250.7.}
\end{figure*}



\label{lastpage}
\bsp	
\end{document}